\documentclass[manuscript=article]{achemso}

\usepackage[version=3]{mhchem} 
\usepackage{color}
\usepackage{multirow}
\usepackage{bigdelim}
\usepackage{booktabs}
\usepackage{nicematrix}
\usepackage{enumitem}
\usepackage{comment}
\usepackage{graphicx}
\usepackage[utf8]{inputenc}
\usepackage[T1]{fontenc}

\SectionNumbersOn

\author{Adam E. A. Fouda}
\affiliation{Chemical Sciences and Engineering Division, Argonne National Laboratory, 9700 S. Cass Avenue, Lemont, IL 60439, USA}
\alsoaffiliation{Department of Physics, The University of Chicago, Chicago, IL 60637, USA}
\email{adamfouda@uchicago.edu}
\author{Bhavnesh Jangid}
\affiliation{Department of Chemistry, The University of Chicago, Chicago, Illinois 60637, United States}
\author{Eetu Pelimanni}
\author{Stephen H. Southworth}
\author{Phay J. Ho}
\affiliation{Chemical Sciences and Engineering Division, Argonne National Laboratory, 9700 S. Cass Avenue, Lemont, IL 60439, USA}
\author{Laura Gagliardi}
\affiliation{Department of Chemistry, The University of Chicago, Chicago, Illinois 60637, United States}
\alsoaffiliation{Pritzker School of Molecular Engineering, The University of Chicago, Chicago, Illinois 60637, United States}
\author{Linda Young}
\affiliation{Chemical Sciences and Engineering Division, Argonne National Laboratory, 9700 S. Cass Avenue, Lemont, IL 60439, USA}
\alsoaffiliation{Department of Physics and James Franck Institute, The University of Chicago, Chicago, Illinois 60637, USA}

\title[An \textsf{achemso} demo]
  {Computation of Auger Electron Spectra in Organic Molecules with Multiconfiguration Pair-Density Functional Theory}

\begin{document}

\begin{abstract}
Efficiently and accurately computing molecular Auger electron spectra for larger systems is limited by the increasing complexity of the scaling in the number of doubly-ionized final states with respect to the system size. In this work, we benchmark the application of multiconfiguration pair-density functional theory with a restricted active space (RAS) reference wave function, for computing the carbon K-edge decay spectra of 21 organic molecules, with decay rates computed within the one-center approximation. The performance of different basis sets and on-top functionals is evaluated and the results show that multiconfiguration pair-density functional theory is comparable in accuracy to RAS followed by second-order perturbation theory, but at a significantly reduced cost and both methods demonstrate good agreement with experiment.

\end{abstract}

\section{Introduction}


The Auger effect, also known as the Auger-Meitner effect, is the non-radiative decay process following inner-shell ionization/excitation. An outer-shell electron fills the core vacancy, and another outer-shell electron is ejected to the continuum. The kinetic energy spectrum of Auger electrons encodes detailed information about the excited atomic site and its local bonding environment, and thereby Auger electron spectroscopy (AES) is widely used in materials analysis and in studying fundamental atomic and molecular physics. New types of inner-shell excitation and decay mechanisms continue to be explored\cite{doi:10.1021/acs.chemrev.0c00106,PhysRevLett.131.253201,Pelimanni2024}, and insights of the underlying ultrafast dynamics are obtained from time-resolved studies at the attosecond timescale\cite{Thompson_2024,D1CP00623A,doi:10.1126/science.abj2096,Driver2024}. In addition, Auger processes play a key role in the formation of radiation damage effects, influencing e.g. structural X-ray diffraction\cite{Stern:dz5149,doi:10.1021/acsnano.7b03447,Carugo2005-lr} and medical radiotherapy applications.\cite{doi:10.1080/09553002.2020.1831706,Ku2019}

The application of AES to increasingly complex systems relies on advancing the accuracy and efficiency of computational methods. Molecular AES calculations are challenging due to both the treatment of the continuum electron wave function, and the number of possible final doubly ionized states, which increases exponentially with the system size. Numerous developments have addressed the treatment of the continuum electron wave function, through either implicit\cite{10.1063/1.1316046,BSchimmelpfennig_1992,SCHIMMELPFENNIG1995173,LIEGENER1982188,10.1063/1.2126976,10.1063/5.0036976,SIEGBAHN1975330,JENNISON1980435,Larkins1990,FINK1995295,TRAVNIKOVA200967,PhysRevA.94.023422,10.1063/1.4919794} or explicit\cite{PhysRevA.19.1649,HIGASHI1982377,PhysRevA.45.318,Demekhin2007,PhysRevA.80.063425,10.1063/1.3526026,10.1063/1.3700233,C7CP02345F} considerations. A number of studies have shown that excluding the continuum and approximating the spectral intensity by an electron population analysis is an effective and low-cost approach for interpreting the spectra\cite{MITANI2003103,10.1063/1.2166234,D3CP01746J,doi:10.1080/00268976.2022.2133749,doi:10.1021/acs.jpclett.3c03611}. Addressing the multitude of final doubly ionized states in AES requires efficient treatment of both the continuum wave function and the bound-state electronic structure of inner-valence cation states. The multiconfigurational nature of these states has been recognized\cite{doi:10.1063/1.1386414}, and recent developments using a multireference description for the bound-state electron structure in the Auger-Meitner decay rate calculation have used the restricted active space self-consistent field (RASSCF) method\cite{werner1981quadratically,malmqvist1990restricted}. RASSCF has been widely used to treat core-hole states\cite{https://doi.org/10.1002/wcms.1433,https://doi.org/10.1002/jcc.24237,doi:10.1021/jz301479j,10.1063/1.4928511,C9CP03019K,doi:10.1021/acs.jpclett.0c01981,Koulentianos_2020,10.1063/5.0050891} and was used in the development of the spherical continuum for ionization (SCI) model by Grell \textit{et al.}\cite{PhysRevA.100.042512,10.1063/1.5142251}. The SCI method explicitly treats the continuum electron wave function to calculate partial Auger-Meitner decay rates by numerically solving the radial Schr\"odinger equation in a spherically averaged potential of the bound-state cation. However, the numerical solution to the continuum wave function is computationally demanding; a recent multireference implementation of the one-center approximation\cite{SIEGBAHN1975330,JENNISON1980435,Larkins1990,FINK1995295,TRAVNIKOVA200967,PhysRevA.94.023422,10.1063/1.4919794} (OCA) by Tenorio \textit{et al.}, implicitly treats the continuum with precalculated bound-continuum integrals from atomic calculations to demonstrate comparable accuracy to SCI at a significantly reduced computational cost\cite{cabraljctc2022}.

In order to extend the application of the multireference OCA implementation \cite{cabraljctc2022} to more complex systems, an efficient description of the large number of final doubly ionized final states is required. Previously, the calculation of the RASSCF wave function was followed by the second order perturbation theory (RASPT2) to include additional effects of the correlation energy\cite{malmqvist2008restricted}, which is typically required to accurately interpret experiments. However, the RASPT2 method requires the calculation of higher-order density matrices to determine the perturbed wave function used to compute the energy. Consequently, the computational cost of this method becomes prohibitive when increasing both the system size and the active space. An alternative approach, recently developed to incorporate additional correlation effects to RASSCF, is multiconfiguration pair-density functional theory\cite{mcpdft2014,mcpdftreview01, mcpdftreview02, mcpdftreview03} (MC-PDFT). MC-PDFT offers a significantly reduced computational cost compared to RASPT2, as the energy is straightforwardly computed using the one- and two-particle reduced density matrices, along with the optimized orbitals derived from the reference multiconfiguration wave function (see Section \ref{sec:theory} for more theoretical details). MC-PDFT has been shown to achieve an accuracy comparable to RASPT2 (or CASPT2) for the energies and properties of the ground and excited states\cite{mcpdftreview02, mcpdftreview03}. Moreover, it is immune to the ``intruder state" problem experienced by RASPT2, where configurations weakly coupled to the state of interest cause singularities complicating the energy computation. MC-PDFT is therefore a robust and efficient alternative to PT2 methods for describing multireference states for spectroscopic applications. MC-PDFT has been mainly used to compute excitation energies\cite{ghosh2023combined, doi:10.1021/acs.chemrev.8b00193}.

\begin{figure}[ht!]
 \centering
 \includegraphics[width=10cm]{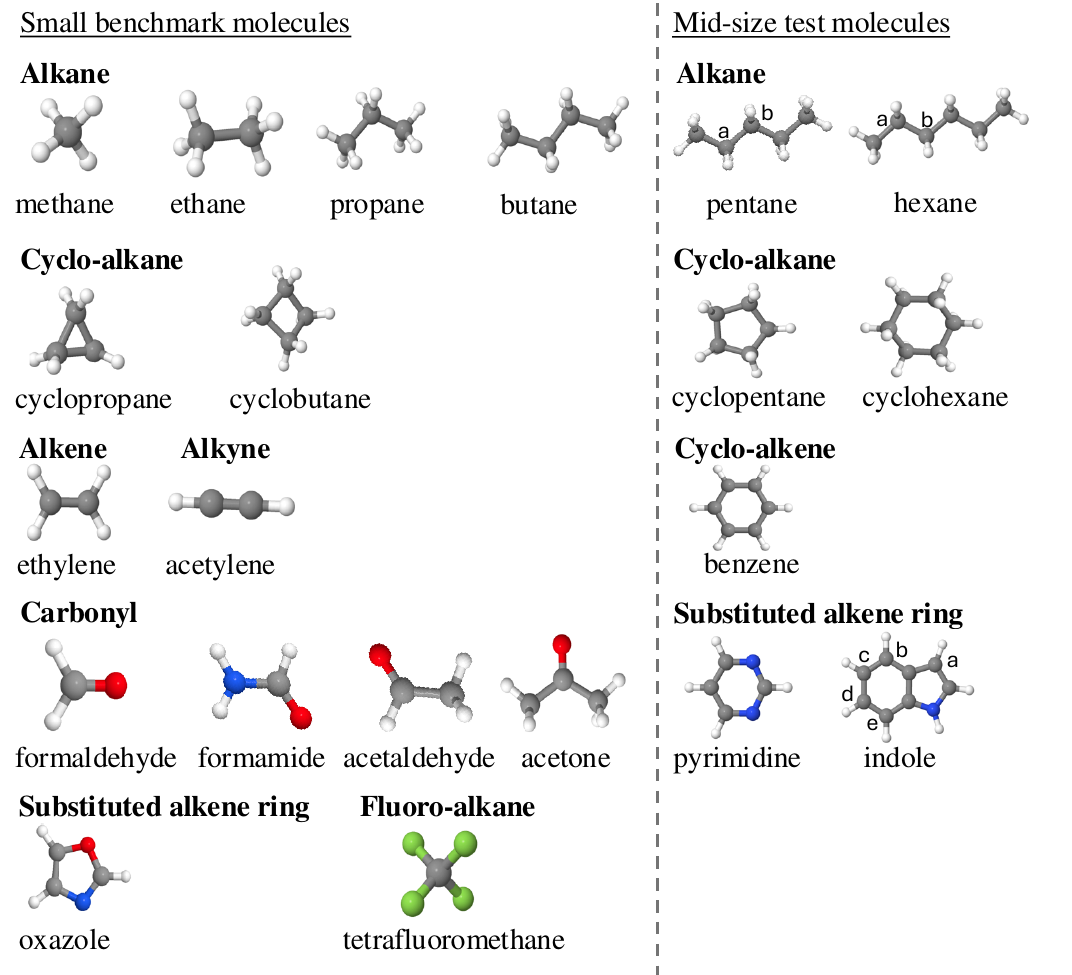}
 \caption{Molecules used in this study.}
 \label{fgr:molecules}
\end{figure}

In this work, we employ MC-PDFT in combination with a general RASSCF active space scheme and OCA decay rates to compute the carbon K-edge Auger electron spectra of various organic molecules (Fig. 1). A set of 14 small molecules, shown on the left side of Fig. 1, is first used to evaluate the performance of MC-PDFT against RASPT2, and to benchmark different basis sets and on-top functionals for MC-PDFT. Then, a set of seven larger, mid-size molecules, shown on the right side of Fig. 1, is used to further test the chosen MC-PDFT method, and the requirement for a multireference description of the final dication states is emphasized by applying the M-value multireference character diagnostic.\cite{doi:10.1021/ct800077r} We find results with comparable accuracy to RASPT2 and in good agreement with experimental references taken from the literature, demonstrating that the MC-PDFT method is an efficient multireference approach for performing AES computations on organic molecules. The paper is organized as follows. In Section 2, we give a theoretical summary of the RASSCF, RASPT2, MC-PDFT and OCA methods. Section 3 presents the computational details and describes a generalized RASSCF active space scheme for non-resonant AES in systems with closed-shell ground states. Section 4 presents results and discussion and Section 5 presents concluding remarks.

\section{Theory}\label{sec:theory}
This section provides a brief overview of the theoretical methods used in this study.\\

\textbf{Restricted Active Space}: The RASSCF method is a multiconfigurational self-consistent field (MCSCF) method routinely used to extend the active space sizes beyond those accessible by the conventional complete active space self-consistent field (CASSCF).\cite{olsen1988determinant,malmqvist1990restricted} Additionally, it is commonly applied to model valence\cite{sauri2011multiconfigurational} and core-excited states\cite{https://doi.org/10.1002/wcms.1433,https://doi.org/10.1002/jcc.24237,doi:10.1021/jz301479j,10.1063/1.4928511,C9CP03019K,doi:10.1021/acs.jpclett.0c01981,Koulentianos_2020,10.1063/5.0050891} as well as charge transfer states,\cite{casanova2022restricted} due to its capability to generate specific, \textit{non-aufbau}, electronic configurations. In the RASSCF framework, the active space is divided into three subspaces: RAS1, RAS2, and RAS3, denoted as ($n_{1}$,$n_{2}$,$n_{3}$; $m_{1}$,$m_{2}$,$m_{3}$).
\begin{itemize}
    \item RAS1: Contains up to $n_{1}$ holes in $m_{1}$ doubly occupied orbitals.
    \item RAS2: Equivalent to the conventional ($n_{2}$,$m_{2}$) CASSCF space, allowing all possible configurations generated by distributing $n_{2}$ electrons across $m_{2}$ orbitals.
    \item RAS3: Contains up to $n_{3}$ particles in $m_{3}$ virtual orbitals.
\end{itemize}

These restrictions on the number of excitations within specific subspaces significantly reduce the number of configurations and, thus the computational cost. This enables the RASSCF to handle larger active spaces than traditional CASSCF approaches.
 
For modeling AES, the intermediate core-ionized state is obtained by placing the core-hole orbital in the RAS1 subspace with one hole, while the remaining non-core occupied orbitals are included in the RAS2 subspace. The active space can be summarized as: (1, $n_{2}$, 0; 1, $m_{2}$, 0). To target highly excited core-ionized states, the core-valence separation (CVS) approximation is used, analogous to the HEXS scheme \cite{delcey2019efficient}. For the final dicationic state, the same number of RAS1 and RAS2 orbitals are used, with the active space represented as (0, $n_{2}-2$, 0; 1, $m_{2}$, 0). Explicit orbital relaxation  is incorporated for both the intermediate and final states by optimizing the orbitals using the Super-CI scheme of OpenMolcas version 24.02\cite{fdez2019openmolcas}.



\textbf{Restricted Active Space Perturbation Theory}: The RASPT2 method\cite{malmqvist2008restricted} incorporates second-order perturbative corrections to the RASSCF energy ($E^{RASSCF}$) to account for dynamic correlation. The RASSCF wave function serves as the reference wave function ($|\Psi^{0}\rangle$), and the RASPT2 energy ($E^{RASPT2}$) is computed by expanding the wave function to first order ($|\Psi^{1}\rangle$) using a double excitation operator.  This introduces the perturbative corrections via excitations between the inactive, active (RAS1, RAS2, and RAS3) and virtual orbital spaces. Here, $\hat{E}_{p}^{q}$ is the spin-preserving excitation operator and $T_{pq}^{rs}$ are the amplitudes.\cite{pulay2011perspective}
\begin{equation}
    |\Psi^{1}\rangle = \sum_{pqrs}T_{pq}^{rs}\hat{E}_{p}^{q}\hat{E}_{r}^{s}|\Psi^{0}\rangle
\end{equation}

The RASPT2 energy is then determined by minimizing the Hylleraas functional
\begin{equation}
    E^{RASPT2} = \min\left(2\langle\Psi^{1}|\hat{H}|\Psi^{0}\rangle + \langle\Psi^{1}|\hat{H}^{0}-E^{0}+E^{SHIFT}|\Psi^{1}\rangle\right)
\end{equation}

While targeting the excited state, RASPT2 often suffers from the ``intruder state problem," where weakly coupled configurations cause singularities in the term \((H_0 - E_0)\), leading to energy divergence. To address this, a level shift parameter ($E^{SHIFT}$) is introduced, incorporating both the real\cite{roos1995multiconfigurational} and imaginary shifts \cite{forsberg1997multiconfiguration}.  These shifts often require careful tuning to ensure convergence. The RASPT2 method, like the CASPT2 method, uses the generalized Fock operator to construct the zeroth-order Hamiltonian, which can result in unbalanced treatment of open- and closed-shell configurations. This introduces systematic errors in excitation energies, bond energies, and spectroscopic constants. The IPEA shift \cite{ghigo2004modified} is applied to the zeroth-order Hamiltonian to mitigate this imbalance, but its utility remains debated \cite{zobel2017ipea}. 

Despite these shortcomings, RASPT2 provides a quite accurate description of the ground and excited states. However, its computational cost rises significantly with larger active spaces and system sizes due to the need to compute third- and higher-order reduced density matrices, intermediate quantities, and integral transformations. Reduced-scale versions of RASPT2 that leverage the sparsity of the orbital space\cite{guo2016sparsemaps, jangid2024efficient} are not yet available.

\textbf{Multiconfiguration Pair-Density Functional Theory}: MC-PDFT extends the Kohn-Sham formalism to strongly correlated systems\cite{mcpdft2014,mcpdftreview01,mcpdftreview02, mcpdftreview03}. The MC-PDFT energy ($E^{MC-PDFT}$) is a function of one-particle ($D_{pq}$) and two-particle ($D_{pqst}$) reduced density matrices, which can be easily obtained via the reference CASSCF/RASSCF wave function.

\begin{equation}
    E^{MC-PDFT} = V_{NN} + \sum_{pq} h_{pq} D_{pq} + \frac{1}{2}\sum_{pqrs} g_{pqrs} D_{pq}D_{rs} + E_{OT}[\rho(r),\Pi(r)] 
\end{equation}

Here, $V_{NN}$ represents the nuclear repulsion energy, $h_{pq}$ includes one-electron integrals (kinetic energy and nuclear-electron attraction), and $g_{pqrs}$ represents two-electron integrals. The on-top density functional, $E_{OT}$, is used to obtain the exchange and correlation energy. Note that the MC-PDFT energy can also be thought as the sum of classical ($E_{Classical}$) and non-classical components ($E^{OT}_{XC}$). The first three terms in the above expression represent the classical energy, while the energy derived from $E_{OT}$ constitutes the non-classical energy. 

$E_{OT}$ depends on both the electron density $\rho(r)$, the on-top pair density $\Pi(r)$, and their derivatives, which can be expressed in terms of \( D_{pq} \), \( D_{pqst} \) and optimized molecular orbitals $\phi(r)$ as follows:
\begin{equation}
\rho(r) = \rho_{\alpha}(r) + \rho_{\beta}(r)= \sum_{pq} \phi_{p}(r)\phi_{q}(r)D_{pq}
\end{equation}
\begin{equation}
\Pi(r) = \frac{1}{2}\sum_{pqst} \phi_{p}(r)\phi_{q}(r)\phi_{s}(r)\phi_{t}(r)D_{pqst}
\end{equation}
The on-top functional is obtained by translating a corresponding KS functional. The translated on-top functional\cite{mcpdft2014}, $E_{OT}[\rho(r),\Pi(r), \rho'(r)]$, are represented as tPBE, tBLYP, etc., while the fully translated on-top functionals\cite{carlson2015multiconfiguration}, $E_{OT}[\rho(r),\Pi(r), \rho'(r), \Pi'(r)]$, are denoted as ftPBE, ftBLYP, and so forth.
The one-to-one correspondence can be shown as
\begin{equation}
E_{OT}[\rho(r),\Pi(r),\rho'(r),\Pi'(r)] = E_{XC}[\tilde{\rho}(r),\tilde{\rho}'(r)] 
\end{equation}

The effective density ($\Tilde{\rho}(r)$) and its gradient ($\Tilde{\rho}'(r)$) are obtained as follows 

\begin{equation}
\Tilde{\rho}(r) = 
\begin{cases}
    \rho(r)(1 \pm \zeta(r)) & \text{if } R(r) \leq 1 \\
    \rho(r) & R(r) > 1
\end{cases}
\quad \text{and} \quad
\Tilde{\rho}'(r) = 
\begin{cases}
    \rho'(r)(1 \pm \zeta(r)) & \text{if } R(r) \leq 1 \\
    \rho'(r) & R(r) > 1
\end{cases}
\end{equation}
Here, $R$ is a function of 3D real space coordinate $r$.
\begin{equation}
    \zeta(r) = \sqrt{1-R(r)} \\ \text{ and } \\
     R(r) = \frac{4\Pi(r)}{[\rho(r)]^{2}}
\end{equation}

By analogy with the``global hybrid" functional in KS formalism, the hybrid on-top functional is defined as a weighted average of the multireference wave function energy and the MC-PDFT energy.\cite{pandharkar2020new, mostafanejad2020global} For example, in the case of tPBE0, $\lambda=0.25$

\begin{equation}
    E^{HMC-PDFT} = (1-\lambda)E^{MC-PDFT} + \lambda E^{RASSCF}
\end{equation}

This can also be presented as follows:
\begin{equation}
    E^{HMC-PDFT} = E_{Classical} + \lambda E^{RASSCF}_{XC} + (1-\lambda)E^{OT}_{XC}
\end{equation}

In terms of computational cost, the classical component of MC-PDFT energy incurs no additional cost beyond the multiconfiguration wave function computation, while the non-classical component involves a quadrature calculation, which is negligible compared to a corresponding RASPT2 calculation. Also, MC-PDFT is free from the symmetry dilemma\cite{mcpdft2014} and reduces the self-interaction error\cite{bao2018self} (SIE) of the KS-DFT formalism.

\textbf{Auger-Meitner Decay Rates:} Here we briefly describe the OCA for computing Auger-decay rates\cite{SIEGBAHN1975330,JENNISON1980435,Larkins1990,FINK1995295}. A more in-depth discussion can be found in the article for the OpenMolcas implementation used in this work\cite{cabraljctc2022}. The decay rates ($\Gamma_{FI;Elm}$) are calculated within the Wentzel's ansatz, which decouples decay of core-hole intermediate from the initial core-ionization process via the formula below,
\begin{equation}
    \Gamma_{FI;Elm} = 2\pi|\langle\Psi^{(-)}_{F;Elm}|\hat{H}-E_{I}|\Psi_{I}\rangle|^{2}, 
\end{equation}
where $\Psi_{I}$ is the core-ionized intermediate state with $N_{I}$ electrons and $\Psi^{(-)}_{F;Elm}$ is the final state.  We approximate this final state by an antisymmetrized product of the bound electronic state $\Psi^{N_{I}-1}_{F}$ and a single-electron continuum wave function $\phi^{(-)}_{\overrightarrow{k}}$. The latter has momentum $\overrightarrow{k}$ and is described by a sum of angular momentum eigenstates in the partial waves approximation, 
\begin{equation}
    \phi^{(-)}_{\overrightarrow{k}} = \sum_{lm}C_{lm\overrightarrow{k}}\phi^{(-)}_{Elm},\quad E = \frac{k^2}{2},
\end{equation}
where the analytical coefficients $C_{lm\overrightarrow{k}}$ contain the spherical harmonics and Coulomb phase contributions. The solution of $\Gamma_{FI;Elm}$ can then be further simplified by considering strong orthogonality between the continuum and bound orbitals to give,
\begin{equation}\label{eq:sorate}
    \Gamma_{FI;Elm} = 2\pi\Big|\sum_{p}\langle\phi_{Elm}|\hat{h}|\phi_{p}\rangle R_{FI;p} + \sum_{qrs}\langle\phi_{Elm}\phi_{q}|\phi_{r}\phi_{s}\rangle R_{FI;qsr} \Big|^2.
\end{equation}
The first term inside the square modulus in Eq. \ref{eq:sorate} is formed from a product of the usual one-electron Hamiltonian operator $\hat{h}$ between the continuum orbital $\phi_{Elm}$ and molecular spin orbitals $\phi_{p}$ and the expansion coefficients of the one-particle Dyson orbital $R_{FI;p}$. This term has been shown to have a small contribution to the decay rate\cite{PhysRevA.100.042512} and this implementation of the OCA further reduces the computation of the rates by only considering the two electron term,
\begin{equation}\label{eq:rate}
  \Gamma_{FI;Elm} \simeq  2\pi\Big| \sum_{qrs}\langle\phi_{Elm}\phi_{q}|\phi_{r}\phi_{s}\rangle R_{FI;qsr} \Big|^2 , 
\end{equation}
which contains a product of the two electron integral involving the continuum orbital $\langle\phi_{Elm}\phi_{q}|\phi_{r}\phi_{s}\rangle$ and the two-particle Dyson matrix $R_{KF;qsr}$. $R_{KF;qsr}$ is available in the OpenMolcas software using the set of biorthonormalized orbitals produced by the restricted active space state interaction (RASSI) method. $\langle\phi_{Elm}\phi_{q}|\phi_{r}\phi_{s}\rangle$ is approximated by a basis of atomic orbitals $\{\chi^{A}_{\lambda}\}$ centered on the core-hole containing atom ($A$),
\begin{equation}
    \langle\phi_{Elm}\phi_{q}|\phi_{r}\phi_{s}\rangle \simeq \sum_{\mu\nu\rho} \langle\chi^{A}_{Elm}\chi^{A}_{\mu}|\chi^{A}_{\nu}\chi^{A}_{\rho}\rangle D_{\mu c}D_{\nu r}D_{\rho s}.
\end{equation}
$D_{\nu r}$ are expansion coefficients defined by projecting the molecular orbitals onto a minimal basis set space, extracted from the standard contracted Gaussian-type orbital basis set and calculated by,
\begin{equation}
    D_{\mu r} = \sum_{\kappa}(T^{-1}U)_{\mu \kappa}C_{\kappa r}.
\end{equation}
$C_{\kappa r}$ are the expansion coefficients of the GTO basis, $T_{\mu\kappa}$ is the overlap matrix of the minimal basis set space, extracted from the GTO basis set, and $U_{\mu\kappa}$ is the overlap between the two basis sets. The atomic two-electron integrals $\langle\chi^{A}_{Elm}\chi^{A}_{\mu}|\chi^{A}_{\nu}\chi^{A}_{\rho}\rangle$ can be numerically solved or extracted from tabulations in  the literature\cite{PhysRev.185.1,WALTERS1971301,CHEN19901}. The OCA implementation in OpenMolcas assumes the high-kinetic energy ($\approx10^{2}$ eV) ejected electrons experience a minor effect from the molecular field and uses a set of energy-independent atomic integrals with the radial part of the integral taken from the literature\cite{PhysRev.185.1}. The angular parts are calculated analytically on the fly\cite{cabraljctc2022}. 

Calculating the Auger-Meitner decay rates by Eq. \ref{eq:rate}, provides the peak heights for the AES discrete stick spectra, which are plotted against the kinetic energies of the ejected electrons ($E_{kinetic}$), calculated by the energy difference of $\Psi_{I}$ and $\Psi^{N_{I}-1}_{F}$, 
\begin{equation}\label{eq:ke}
    E_{kinetic} = E_{I} - E^{N_{I}-1}_{F}.
\end{equation}
A Gaussian broadening is then applied to the discrete spectra. Details of this broadening are provided in the following section.

\section{Computational Details}\label{sec:comp}

The present work uses optimized molecular geometries reported in the NIST WebBook\cite{147901}, with DFT B3LYP and the 6-311G$^{**}$ basis set, except for cyclohexane, hexane, pentane, pyrimidine and tetrafluoromethane which used the 6-311G$^{*}$ basis set. Unless stated otherwise, all electronic structure calculations employed the ANO-RCC-VTZP basis set, incorporating scalar relativistic effects via the second-order Douglas-Kroll-Hess (DKH) Hamiltonian\cite{wolf2002generalized, reiher2004exact, Roos2004}. No point-group symmetry constraints were used in this work and all molecules have closed-shell neutral ground states.

\begin{figure}[ht!]
 \centering
 \includegraphics[width=12cm]{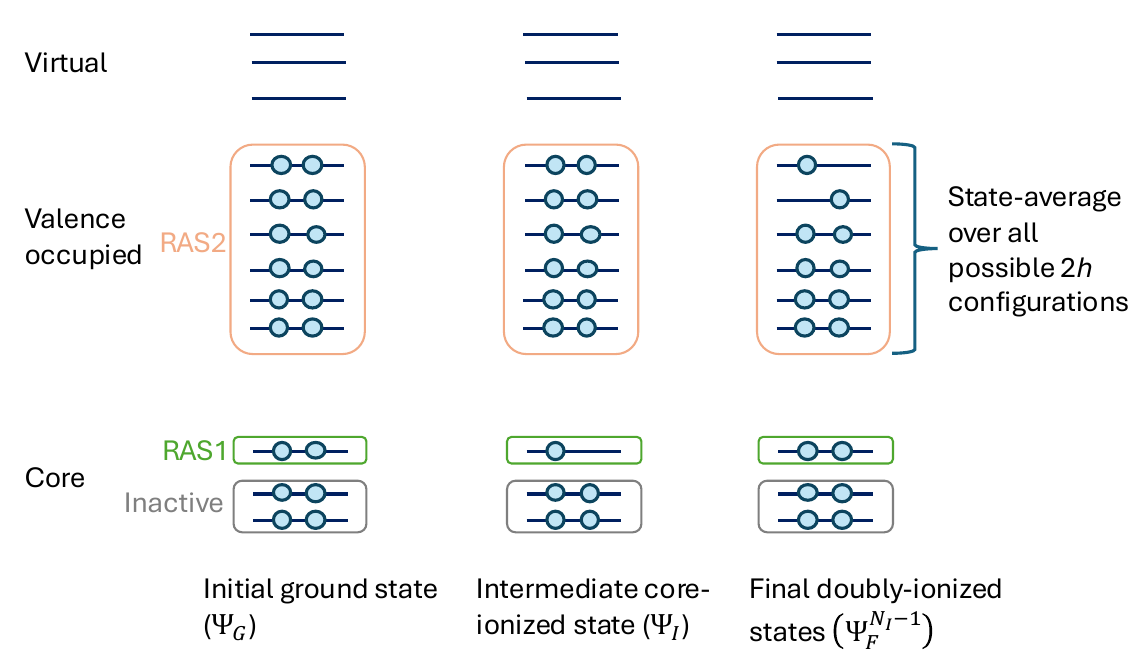}
 \caption{RASSCF active space scheme applied to all molecules in this study. All non-1$s$ occupied orbitals are placed into RAS2. The core-ionized orbital is placed into RAS1, the remaining core orbitals are inactive. RAS3 is not used and virtual orbitals are excluded from the active space.}
 \label{fgr:rasscheme}
\end{figure}

Fig. \ref{fgr:rasscheme} outlines the RASSCF active space scheme used for computing ground, core-ionized, and valence dication states across all molecules. In this scheme, the core-hole orbital is assigned to RAS1 with a maximum hole occupation of 1, while the remaining core orbitals are treated as inactive. Non-1$s$ occupied orbitals are placed in RAS2, and no virtual orbitals are included in RAS3. The scheme is general for non-resonant AES computations in any molecule with closed-shell ground states and does not account for any shake-up/off transitions. Each of the ground, core-ionized and dication state RASSCF calculations uses initial guess orbitals from a preliminary Hartree-Fock (HF) calculation with Pipek-Mezey localization\cite{10.1063/1.456588} applied to all occupied orbitals. Core-ionized states are generated using the highly excited state (HEXS) scheme\cite{delcey2019efficient}, while final dication states are computed by removing two electrons from RAS2, followed by a state-averaged RASSCF calculation over all possible singlet and triplet two-hole states.

Both RASPT2\cite{malmqvist2008restricted} and MC-PDFT\cite{mcpdft2014} calculations were performed using OpenMolcas\cite{fdez2019openmolcas}(version 24.02)  to account for out-of-active-space electron correlation. Unless otherwise noted, MC-PDFT calculations employed the tPBE0 on-top functional. Single-state RASPT2 was utilized with an imaginary shift of 0.01 a.u. and a an IPEA shift of 0.25. The Cholesky decomposition of two-electron integrals (threshold of $10^{-4}$ a.u.) using atomic compact auxiliary bases\cite{aquilante2007unbiased} was used throughout. Auger-Meitner decay rates were computed using the OCA implementation in OpenMolcas\cite{cabraljctc2022}.

For the AES spectra, a quantitative comparison is made with the full experimental traces, with a few key considerations outlined here. Although explicit calculations of vibrational and lifetime broadening effects are omitted, they are accounted for by applying a Gaussian broadening of 1.6 eV to all calculated transitions to enable the quantitative comparison with experimental traces. This value is somewhat arbitrary but representative. It is derived from averaging two mean width values of all individual C 1$s$ transitions reported in previous studies on HCNO\cite{D2CP02104H} and HNCS\cite{D4CP03104K} calculated by moment theory (which uses the lifetime energy width of the core-hole intermediate, the gradients of the intermediate and final states, and the ground state vibrational frequencies for calculating the spectral widths \cite{10.1063/1.464348}). In the supplementary material (SM), we show how varying the Gaussian width impacts agreement with experiments (Fig. S1). Additionally, experimental resolution values (summarized in SM Table S1) were incorporated into the Gaussian broadening, except where resolution data was unavailable. For cycloalkanes, the resolution is 2 eV, which is larger than the lifetime and vibrational width applied here and only the experimental resolution is applied to the width in these cases. Given the broad kinetic energy range (40–50 eV) of the Auger electrons, potential contributions from an inhomogeneous background or non-uniform transmission function of the analyzer may introduce uncertainties not explicitly addressed in the references. Furthermore, photon energies in the experimental studies were typically tens of eV above the core-ionization threshold, minimizing postcollision interaction-induced shifts in CEBEs and Auger lines to within a few tens of meV\cite{van1988angular}. The experimental Auger spectra were converted to numerical data using an image analysis tool\cite{WebPlotDigitizer}.



\section{Results and Discussion}\label{sec:results}

\subsection{Comparison of MC-PDFT and RASPT2 Performance for AES}\label{sec:method}

\begin{figure}[ht!]
 \centering
 \includegraphics[width=17cm]{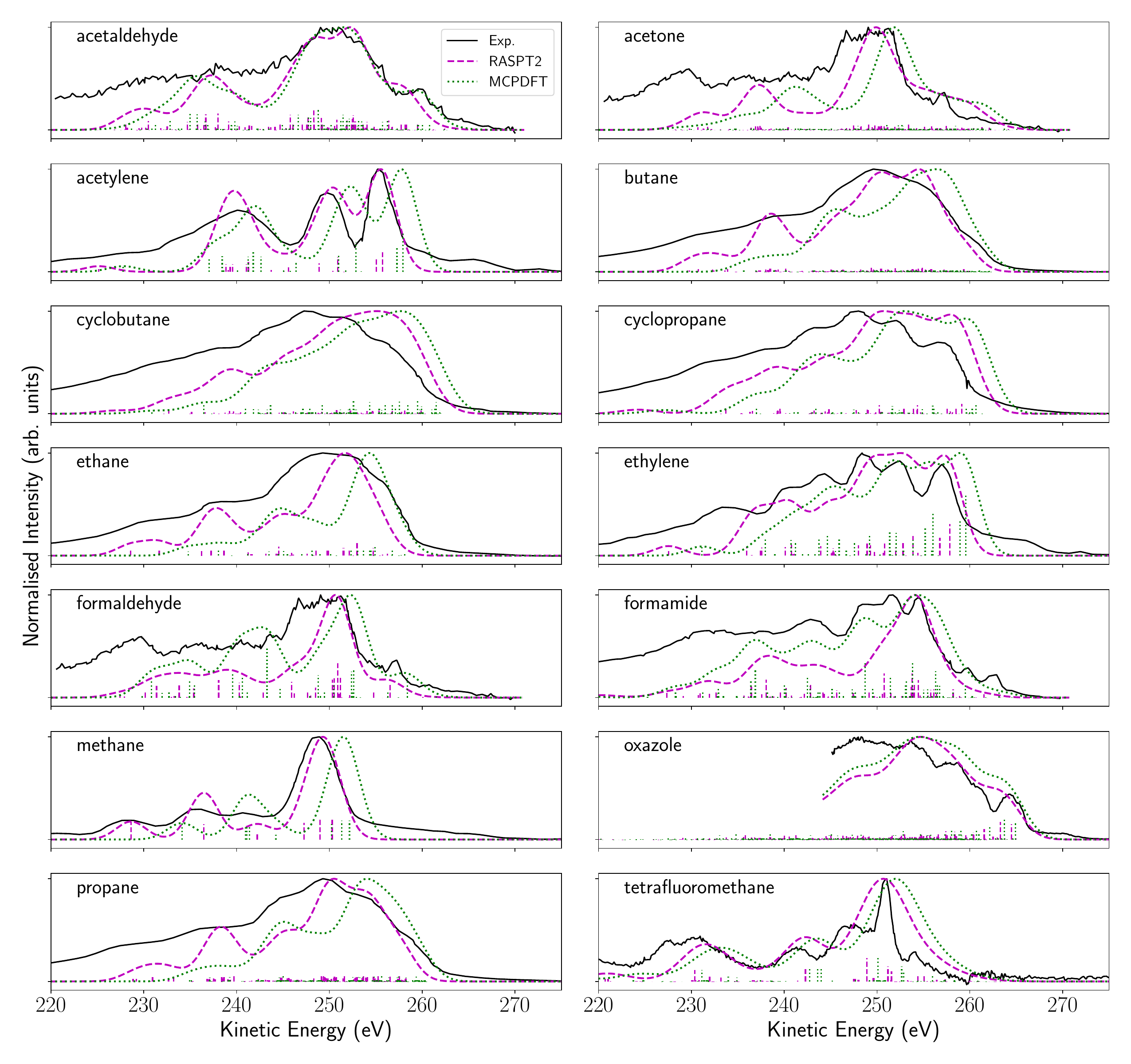}
 \caption{Comparison of RASPT2 and MC-PDFT tPBE0, calculations use the ANO-RCC-VTZP basis set. Experimental spectra are from the following references: acetaldehyde\cite{10.1063/1.461687}, acetone\cite{10.1063/1.461687}, acetylene\cite{10.1063/1.440831}, butane\cite{10.1063/1.440015}, cyclobutane\cite{10.1063/1.440831}, cyclopropane\cite{10.1063/1.440831}, ethane\cite{10.1063/1.440015}, ethylene\cite{10.1063/1.440831}, formaldehyde\cite{10.1063/1.461687}, formamide\cite{10.1063/1.461687}, methane\cite{10.1063/1.440015}, oxazole\cite{10.1063/5.0122088}, propane\cite{10.1063/1.440015}, tetrafluoromethane\cite{MNeeb_1997}.}
 \label{fgr:method}
\end{figure}

First, we compare the performance of MC-PDFT and RASPT2 for computing AES. All calculations in this subsection use the ANO-RCC-VTZP basis set and the MC-PDFT calculations use the tPBE0 on-top functional. Figure \ref{fgr:method} shows RASPT2 and MC-PDFT AES calculations for the 14 benchmark molecules compared to the experimental reference spectra. In order to provide a more quantitative measure of the accuracy, Table \ref{tbl:methodpcc} gives the Pearson's correlation coefficient (PCC) values for each of the calculated spectra. The PCC metric measures the similarity between computed and experimental lineshapes by calculating the cosine similarity between two vectors and their means. A PCC value of 1 indicates perfect agreement. This metric has been previously shown to be reliable for evaluating the accuracy of high-throughput x-ray absorption spectroscopy (XAS) calculations \cite{Suzuki2019} and has recently been applied to validate computed XAS spectra for inferring structural properties with machine learning\cite{doi:10.1021/acs.chemmater.3c02584}. The PCC values are influenced by the width of the Gaussian broadening applied to the transitions and in SM Figure S1 we analyze the effect of varying the Gaussian broadening width the on PCC values. 

Figure \ref{fgr:method} and Table \ref{tbl:methodpcc} show that both RASPT2 and MC-PDFT give good overall agreement with the experimental AES lineshapes, when used with the general RASSCF active space scheme and OCA decay rates. For both RASPT2 and MC-PDFT, the lineshapes and PCC values for acetaldehyde, ethane, ethylene, formamide, oxazole and propane are in strong agreement with experiment. In some cases, such as acetone, acetylene, methane and tetrafluoromethane, the performance of MC-PDFT is worse than RASPT2 and the average PCC value of 0.86 for RASPT2 is higher than 0.75 for MC-PDFT. However, the additional cost of MC-PDFT following the RASSCF calculation is negligible compared to RASPT2. For example, calculating the energies of the 136 singlet dication states of tetrafluormethane on a single thread in OpenMolcas takes 3 hours and 21 minutes with RASPT2 and 16 minutes with MC-PDFT, a factor of 13 times faster. Therefore, results show that MC-PDFT is a reliable approach for molecular AES simulation and suitable for larger systems such as the mid-size test molecules discussed in subsection \ref{sec:large}. We note that as both methods use energy-independent OCA decay rates, the discrepancies in the lineshapes between the methods are due to the energy separations of the individual transitions and not the individual peaks heights. 

\begin{table}[h]
  \caption{Comparison of RASPT2 and MC-PDFT PCC and CEBE MAD (eV) values.}
  \label{tbl:methodpcc}
  \setlength{\tabcolsep}{12pt} 
  \begin{tabular}{@{\extracolsep{\fill}}llllll}
    \hline
      Molecule     & RASPT2 & MC-PDFT  \\
    \hline
    acetaldehyde   & 0.89   & 0.87    \\
    acetone        & 0.71   & 0.52    \\       
    acetylene      & 0.91   & 0.76    \\
    butane         & 0.95   & 0.84    \\
    cyclobutane    & 0.85   & 0.72    \\
    cyclopropane   & 0.87   & 0.76    \\
    ethane         & 0.93   & 0.82    \\
    ethylene       & 0.92   & 0.86    \\  
    formaldehyde   & 0.82   & 0.67    \\    
    formamide      & 0.75   & 0.79    \\   
    methane        & 0.95   & 0.64    \\    
    oxazole        & 0.87   & 0.87    \\
    propane        & 0.92   & 0.82    \\
tetrafluoromethane & 0.74   & 0.61    \\    
    \hline 
    Average        & 0.86   & 0.75    \\
    \hline
  \end{tabular}
\end{table}


The mean average deviations (MAD) for the core-electron binding energy (CEBE) values calculated by RASPT2 and MC-PDFT are 1.63 and 0.27 eV respectively. A previous study evaluated RASPT2 and MC-PDFT for first row $p$-block element CEBEs in small open-shell molecules, finding mean unsigned errors of 1.07 and 0.89 eV respectively\cite{doi:10.1021/acs.jpcc.4c01750}. The active spaces used in the previous study were designed for CEBE values in open-shell systems and therefore include virtual orbitals in the RAS2 space in-order to describe the open-shell ground states. The present study is focused on AES in molecules with a close-shell ground state, and uses the active space scheme shown in Fig. \ref{fgr:rasscheme} which excludes virtual orbitals from either RAS2 or RAS3. Hence the RASSCF CEBE calculations are now effectively a $\Delta$SCF (HF) calculation. In the SM Tables S2 and S3 we present the RASSCF, RASPT2 and MC-PDFT CEBE values for the individual molecules and provide a discussion on how the values are sensitive to the localization procedure applied to the initial guess orbitals for the RASSCF calculation. This highlights the complexity of core-hole state optimization, but the main focus of this work is AES simulation. 

\subsection{Investigation of Basis Sets and On-top Functionals} \label{sec:basis}

\begin{figure}[ht!]
 \centering
 \includegraphics[width=17cm]{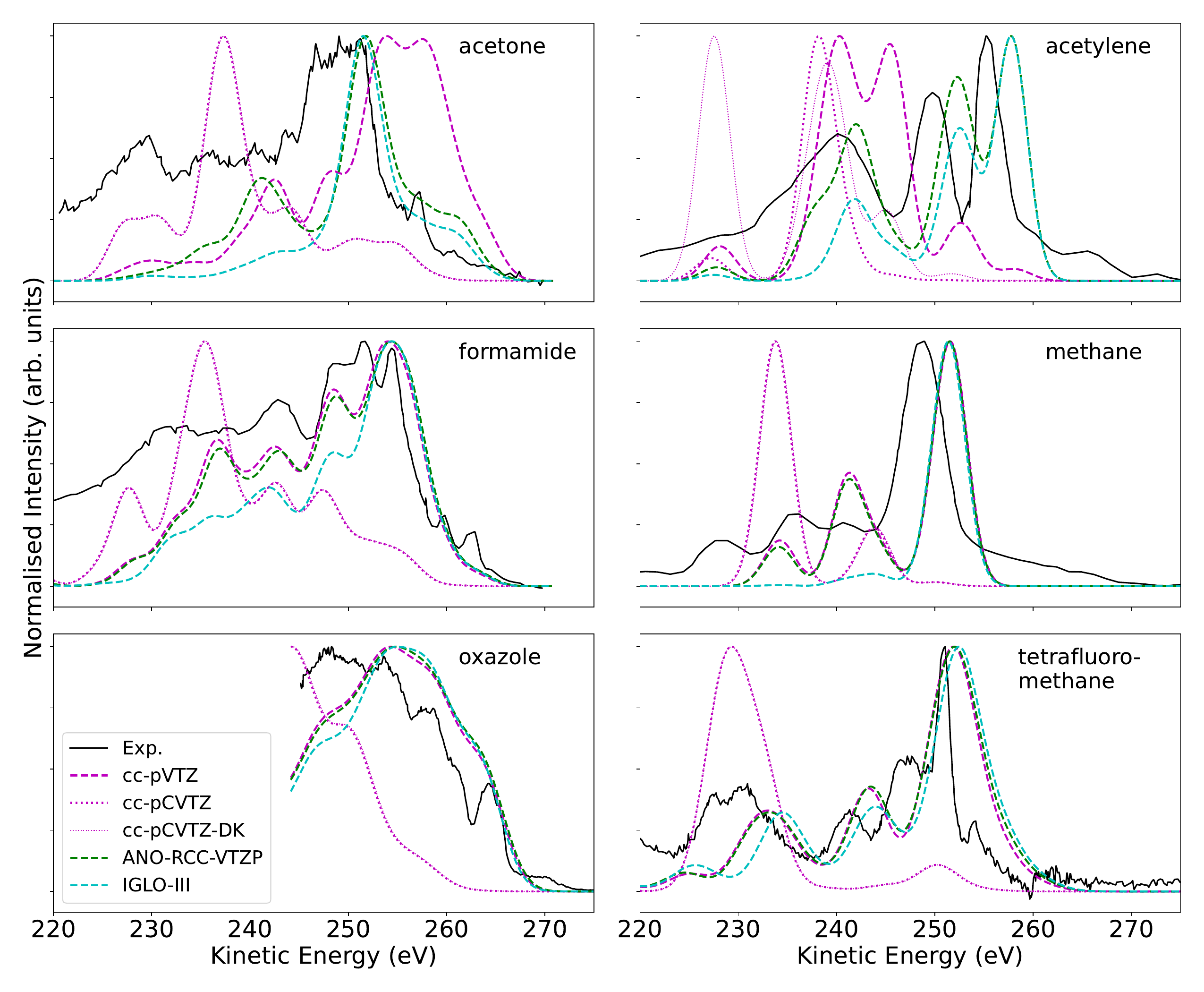}
 \caption{Comparison of different triple zeta basis sets with MC-PDFT. Experimental spectra are from the following references: acetone\cite{10.1063/1.461687}, acetylene\cite{10.1063/1.440831}, formamide\cite{10.1063/1.461687}, methane\cite{10.1063/1.440015}, oxazole\cite{10.1063/5.0122088},  tetrafluoromethane\cite{MNeeb_1997}.}
 \label{fgr:basis}
\end{figure}

\begin{figure}[ht!]
 \centering
 \includegraphics[width=17cm]{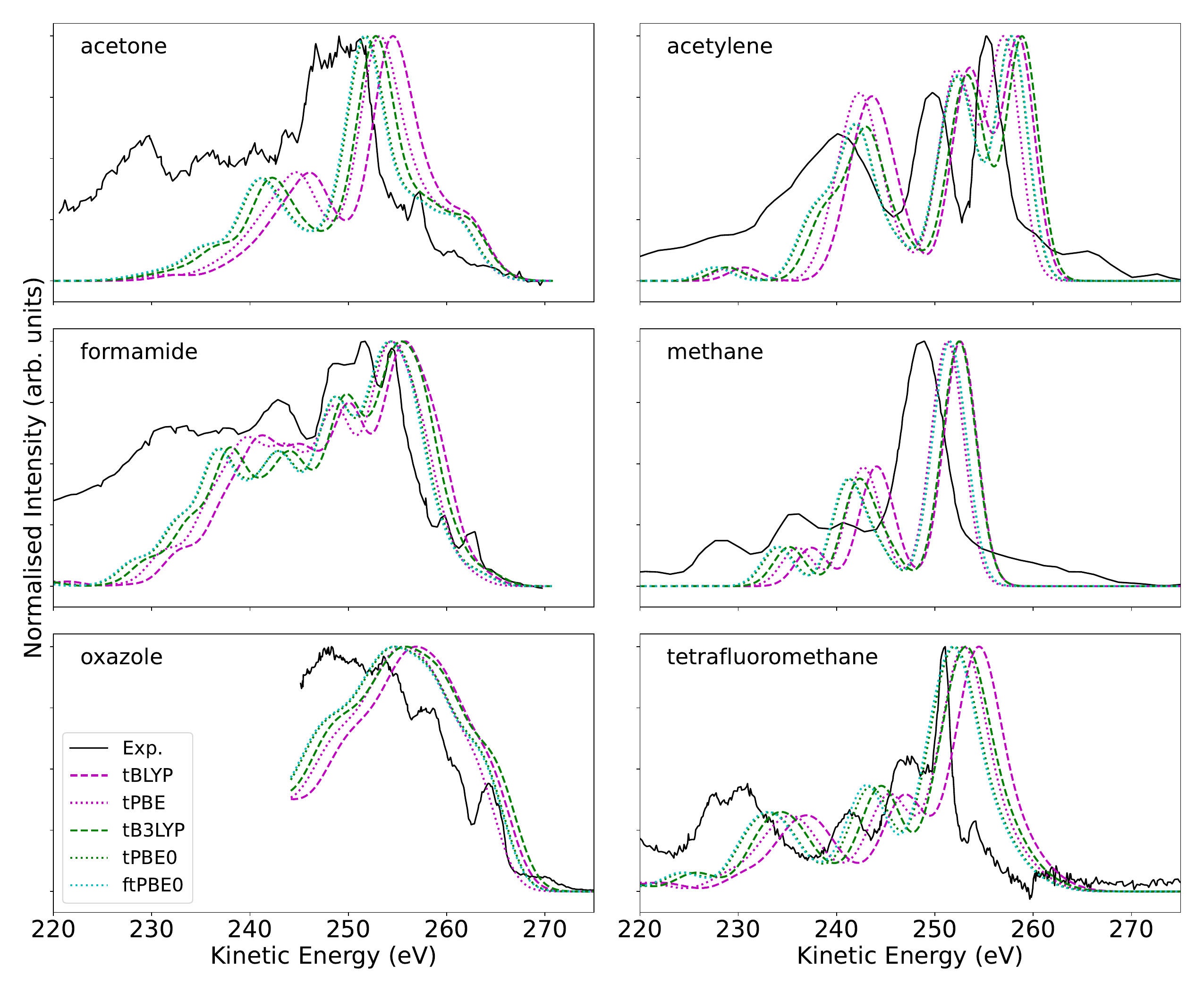}
 \caption{Comparison of different on-top functionals with MC-PDFT. Experimental spectra are from the following references: acetone\cite{10.1063/1.461687}, acetylene\cite{10.1063/1.440831}, formamide\cite{10.1063/1.461687}, methane\cite{10.1063/1.440015}, oxazole\cite{10.1063/5.0122088}, tetrafluoromethane\cite{MNeeb_1997}.}
 \label{fgr:functional}
\end{figure}

Here we present a systematic benchmark of basis sets and on-top functionals that led to the choice of the tPBE0 on-top functional and ANO-RCC-VTZP basis set.  Figures \ref{fgr:basis} and \ref{fgr:functional} present the AES for different basis sets and on-top functionals respectively, compared to experiment for acetone, acetylene, formamide, methane, oxazole and tetrafluoromethane. The results for the full set of molecules is given in SM Figures S2 and S3. Table \ref{tbl:basisfunctional} presents the average AES PCC values and CEBE MAD values for the 14 benchmark molecules. The basis set dependence of the individual PCC and CEBE values is presented in SM Tables S4 and S5 respectively, and the dependence on the on-top functional in SM Tables S6 and S7 respectively. When the basis set is varied the tPBE0 on-top functional is used and when the on-top functional is varied, the ANO-RCC-VTZP basis set is used.

We consider five different triple-zeta basis sets, as triple-zeta or larger is typically required for experimental accuracy. Three correlation consistent basis sets are included: cc-pVTZ (which includes two variations), cc-pCVTZ (which adds core-polarization functions for describing core orbitals) and cc-pCVTZ-DK (a recontracted version of cc-pCVTZ optimized for use with the Douglas-Kroll Hamiltonian to incorporate relativistic effects). We also consider the relativistic core-corrected atomic natural orbital basis set ANO-RCC-VTZP, which is optimized for semi-core electrons and relativistic corrections using the Douglas-Kroll Hamiltonian. Finally, the IGLO-III basis set, originally developed for nuclear magnetic resonance (NMR) shielding constants, is included. IGLO-III provides a detailed description of orbitals near nuclei and has been shown to efficiently and accurately describe core-ionized and -excited states, particularly for hydrogen and first- and second-row p-block elements states\cite{Fouda2017}.

Figure \ref{fgr:basis} shows significant discrepancies in the AES lineshapes between the basis sets. While the ANO-RCC-VTZP and IGLO-III basis sets are consistent with each other and show good agreement with experimental results, the cc basis sets exhibit less consistent performance. For formamide, methane, oxazole and tetrafluoromethane, cc-pVTZ produces lineshapes similar to those of ANO-RCC-VTZP and IGLO-III, showing good agreement with experimental observations. However, for acetone and acetylene, the agreement with experiment is worse, with ANO-RCC-VTZP and IGLO-III outperforming cc-pVTZ. In all cases, cc-pCVTZ and cc-pCVTZ-DK agree with each other, but show significant deviations from the experimental lineshapes. The AES calculation discrepancies can be attributed to differences in the description of the non-1$s$ orbitals and thus the dication final states. This is supported by SM Table S8, which presents the carbon atom contraction for each basis set. The correlation consistent basis sets have a lower number of $p$-type basis functions than ANO-RCC-VTZP and IGLO-III, which are important for describing final states with valence hole orbitals. However cc-pVTZ, contains less $p$-type basis functions than cc-pCVTZ and cc-pCVTZ-DK, and its superior performance is possibly a result of an error cancellation. 

Table \ref{tbl:basisfunctional} shows that all basis sets give sub-eV accuracy for the CEBE predictions. Among them, cc-pVTZ, ANO-RCC-VTZP and IGLO-III show good overall PCC values for AES calculations. ANO-RCC-VTZP, with the highest average PCC value and lowest CEBE MAD value, was the basis set of choice to be used in subsections \ref{sec:method} and \ref{sec:large}. The strong performance of the IGLO-III basis set for both AES and CEBEs suggests potential for future development of computationally efficient basis sets for AES simulations in organic molecules, particularly since IGLO-III lacks $f$- or $g$-type basis functions (see SM Table S8).

\begin{table}[ht]
  \caption{Effect of basis set and on-top functional on MC-PDFT average PCC and CEBE MAD values (eV).}
  \label{tbl:basisfunctional}
  \setlength{\tabcolsep}{12pt} 
  \begin{tabular}{@{\extracolsep{\fill}}llllll}
    \hline
                 & PCC Avg.  & CEBE MAD (eV)\\
    \hline 
    \textbf{Basis set} \\
    cc-pVTZ      & 0.70      & 0.30\\
    cc-pCVTZ     & 0.59      & 0.70\\
    cc-pCVTZ-DK  & 0.58      & 0.58\\
    ANO-RCC-VTZP & 0.75      & 0.27\\
    IGLO-III     & 0.66      & 0.57\\
    \hline        
    \textbf{On-top Functional}    \\
    tBLYP       & 0.59       & 0.71\\
    tPBE        & 0.71       & 0.35\\
    tBLYP0      & 0.67       & 0.54\\
    tPBE0       & 0.75       & 0.27\\ 
    ftPBE0      & 0.76       & 0.28\\
    \hline
  \end{tabular}
\end{table}

For the MC-PDFT functionals, we consider both tBLYP and tPBE, their hybrid variations tBLYP0 and tPBE0, and the fully translated version of tPBE0, ftPBE0. The ftPBE0 functional also includes the gradient of the on-top density and was shown to improve the performance for molecules containing transition metals\cite{doi:10.1021/acs.jctc.5b00609}. Figure \ref{fgr:functional} shows that all on-top functionals provide comparable accuracy for the AES predictions, as reflected by the average PCC values in Table \ref{tbl:basisfunctional}. Figure \ref{fgr:functional} and the PCC values for individual molecules in SM Table S6, show that the hybrid on-top functionals consistently outperform their non-hybrid counterparts. MC-PDFT on-top functional hybridization differs from Kohn-Sham DFT (KSDFT), where a fraction of the HF energy expression is combined with the KSDFT energy to correct for self-interaction errors in the exchange-correlation functional. In MC-PDFT, hybridization involves combining a portion of the MCSCF (or RASSCF, in our case) energy with the MC-PDFT energy expression to correct for the nonclassical component of the energy. This approach has been shown to reduce errors in excitation energies\cite{pandharkar2020new}, and appears to improve the accuracy of Auger electron kinetic energies in our case. We find that tPBE0 and ftPBE0 give the best performance for the CEBE values. Therefore, tPBE0 in combination with the ANO-RCC-VTZP basis set (tPBE0/ANO-RCC-VTZP) was used for subsequent calculations.

\subsection{Application to Larger Molecules and Individual Carbon Environments}\label{sec:large}

\begin{figure}[ht!]
 \centering
 \includegraphics[width=8.5cm]{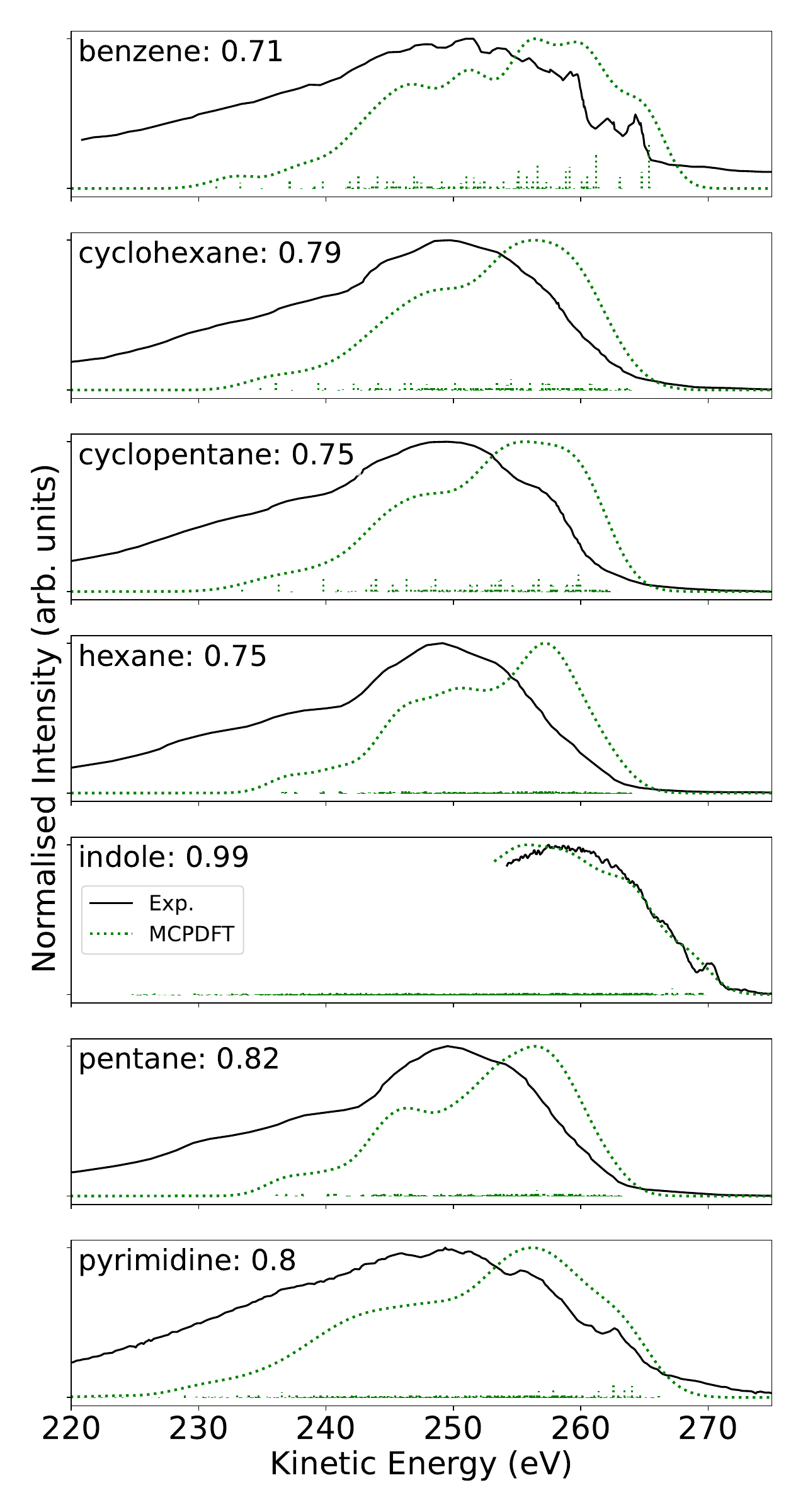}
 \caption{MC-PDFT tPBE0 ANO-RCC-VTZP Auger electron spectra on larger molecules. PCC values are given with the names of the molecules. Experimental spectra are from the following references: benzene\cite{10.1063/1.1290029}, cyclohexane\cite{10.1063/1.451719}, cyclopentane\cite{10.1063/1.451719}, hexane\cite{PhysRevA.14.2133}, indole\cite{doi:10.1021/acs.jpca.0c02719}, pentane\cite{PhysRevA.14.2133}, pyrimidine\cite{10.1063/1.2993317}}
 \label{fgr:large}
\end{figure}

\begin{figure}[ht!]
 \centering
 \includegraphics[width=17cm]{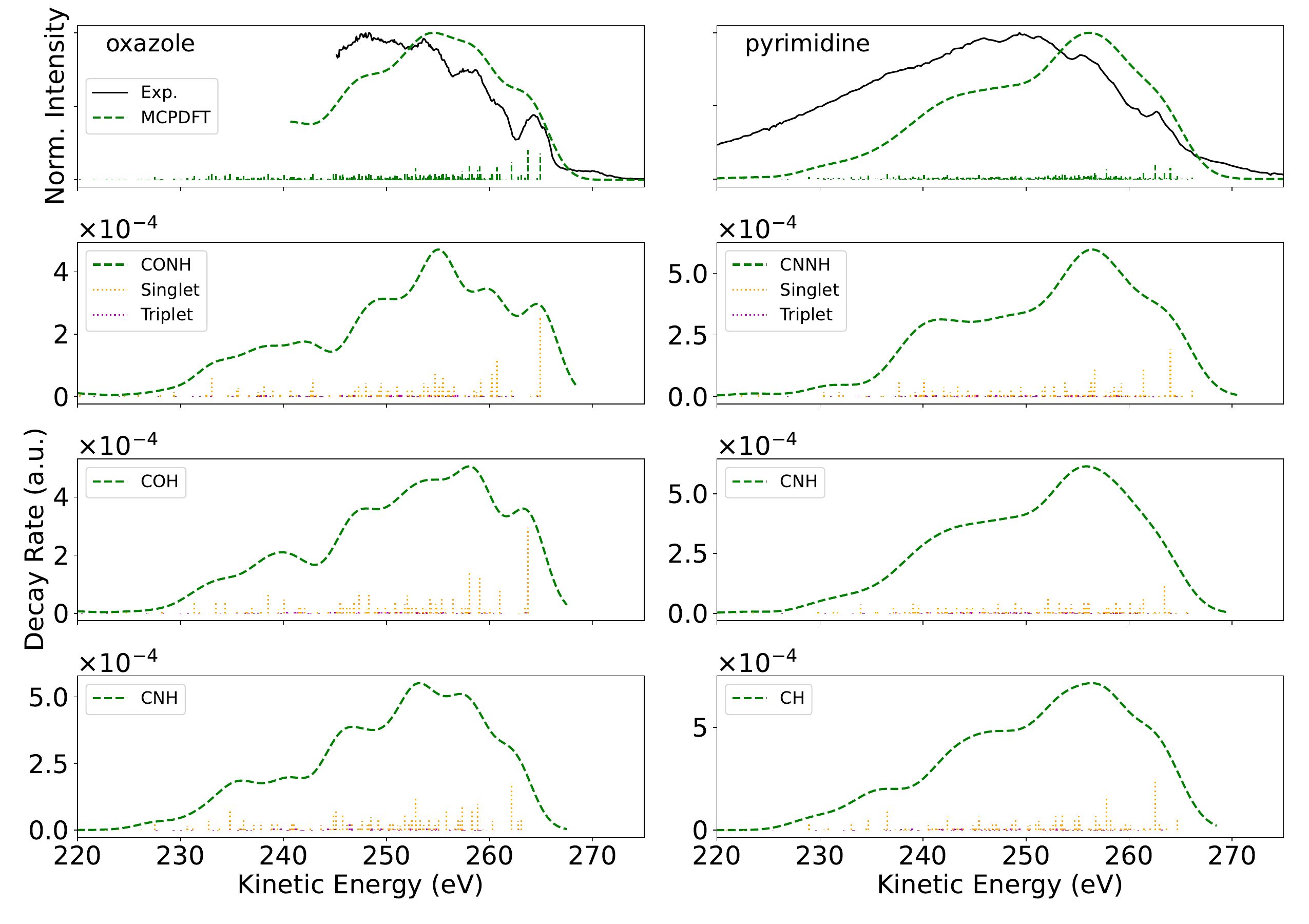}
 \caption{Individual carbon contributions to the spectra of oxazole and pyrimidine calculated by MC-PDFT tPBE0 ANO-RCC-VTZP. Experimental spectra are from the following references: oxazole\cite{10.1063/5.0122088} , pyrimidine\cite{10.1063/1.2993317}.}
 \label{fgr:carbon_all}
\end{figure}

We employed tPBE0/NO-RCC-VTZP for the set of seven larger molecules. Figure \ref{fgr:large} shows the spectra with the PCC values. SM Table S9 reports the CEBEs, which are all in excellent agreement with experiment and the CEBE MAD is 0.28 eV. Notably, the experimental spectra for the larger molecules exhibit broader and less distinct features compared to the smaller molecules, due to contributions from a larger manifold of final states.  Figure \ref{fgr:carbon_all} examines the contributions of individual carbon environments to the overall spectra for oxazole (left) and pyrimidine (right). The top panels show the full spectra, while the three lower panels show the contributions from individual carbon environments with the individual transitions to singlet and triplet final dication states represented as orange and magenta sticks respectively. The carbon environments are labeled by ``C" followed by the nearest non-carbon bonding neighbors and are ordered by decreasing electronegativity. Although the shapes of the spectra vary between environments, no clear relationship between spectral lineshape and local bonding environment was observed. Future studies could leverage larger AES datasets and employ featurization or dimensionality reduction techniques\cite{doi:10.1021/acs.chemmater.3c02584} to uncover such relationships. This could enhance the analytical capabilities of site-specific AES experiments.

Table \ref{tbl:mvalue} presents an analysis of the multiconfigurational character of the dication final states for these systems using the M-value diagnostic\cite{doi:10.1021/ct800077r}. The M-value, derived from the natural orbital occupations of the RASSCF dication final states, quantifies the degree of multiconfigurational character, with M = 0 for a Hartree-Fock wavefunction and larger values indicating higher static correlation. In Table \ref{tbl:mvalue} we give the number of final states, the Mean M-value, the M-value range, the number of multireference (MR) states (M $\geq$ 0.05) and the number of single-reference (SR) states (M $<$ 0.05) for the singlet and triplet dication final states of the 7 larger molecules considered in this work. All molecules exhibit a majority of multireference (MR) final states, with singlet states consistently showing a higher proportion of MR character. The maximum M-value for open-shell systems is slightly above unity, while the mean and maximum M-values of approximately 0.35 and 0.7, respectively, indicate a high degree of multiconfigurational character in the dication final states. These findings emphasize the necessity of multiconfigurational quantum chemistry methods for accurately modeling AES in complex systems. 

\begin{table}[h]
  \caption{M-value diagnostic analysis for the RASSCF wavefunctions of the dication final states of the larger molecules. M values greater than 0.05 are used to define multi-reference (MR) states, less than or equal to 0.05 defines single-reference (SR) states.  }
  \label{tbl:mvalue}
  \setlength{\tabcolsep}{6pt} 
  \begin{NiceTabular}{@{\extracolsep{\fill}}lllllll}
    \hline
      Molecule     &  Spin     &  No. States & Mean M-Value & M-value Range & No. MR & No. SR \\
    \hline
    benzene        &  Singlet  &  120        & 0.336        & 0.004-0.750   & 117           & 3   \\
                   &  Triplet  &  105        & 0.331        & 0.000-0.750   & 100           & 5   \\      
    \hline
    cyclohexane    & Singlet   &  171        & 0.317        & 0.120-0.706   & 171           & 0    \\
                   & Triplet   &  153        & 0.320        & 0.037-0.729   & 151           & 2    \\         
    \hline        
    cyclopentane   & Singlet   &  120        & 0.383        & 0.104-0.703   & 120           & 0    \\
                   & Triplet   &  105        & 0.374        & 0.023-0.707   & 101           & 4    \\ 
    \hline 
    hexane         & Singlet   &  190        & 0.348        & 0.014-0.716   & 189           & 1    \\
                   & Triplet   &  171        & 0.342        & 0.012-0.703   & 168           & 3    \\ 
    \hline
    indole         & Singlet   &  253        & 0.335        & 0.014-0.683   & 252           & 1    \\
                   & Triplet   &  231        & 0.330        & 0.003-0.682   & 218           & 13   \\ 
    \hline
    pentane        & Singlet   &  136        & 0.360        & 0.119-0.679   & 136           & 0    \\
                   & Triplet   &  120        & 0.350        & 0.021-0.744   & 116           & 4    \\ 
    \hline
    pyrimidine     & Singlet   &  120        & 0.383        & 0.001-0.734   & 118           & 2    \\
                   & Triplet   &  105        & 0.405        & 0.000-0.734   & 97            & 8    \\ 
    \hline
  \end{NiceTabular}
\end{table}

\section{Conclusion}

MC-PDFT, combined with the additional efficiencies introduced by the general RAS scheme and the OCA decay rates, provides a computationally viable approach to account for large numbers of multiconfigurational final dication states inherent to AES computations of organic molecules. Whilst the agreement with experiment is comparable to RASPT2, the additional cost of MC-PDFT to the RASSCF calculation is negligible in comparison to the additional cost of RASPT2. 
Therefore, the presented approach is well suited for AES simulation on larger molecules where the number of possible final states, increases exponentially with the system size. 

The novel general active space scheme presented by this work, does not require specific active space choices for each systems and could therefore automate the simulation of AES for future high-throughput investigations. Future data-driven studies of AES could employ feature extraction techniques to analyze spectral data\cite{doi:10.1021/acs.chemmater.3c02584}, potentially uncovering relationships between spectral lineshapes and local bonding environments to yield more quantitative structural insights. Another promising direction is integrating this approach with quantum embedding methods,\cite{jangid2024core} which can selectively apply more accurate techniques to specific regions of a system whilst treating the surrounding environment with a lower level of theory. 



\section*{Acknowledgment}
This work was supported by the U.S. Department of Energy, Office of Basic Energy Sciences, Division of Chemical Sciences, Geosciences, and Biosciences through Argonne National Laboratory. 
This work was also partially
supported (BJ and LG) by the Computational Chemical Sciences
Program, under Award DE-SC0023382, funded by the U.S.
521 Department of Energy, Office of Basic Energy Sciences,
522 Chemical Sciences, Geosciences, and Biosciences Division.
Argonne is a U.S. Department of Energy laboratory managed by UChicago Argonne, LLC, under contract DE-AC02-06CH11357. A. E. A. F. is grateful for the support from the Eric and Wendy Schmidt AI in Science Postdoctoral Fellowship, a Schmidt Futures Program.
The computational resources were provided by the Research Computing Center at the University of Chicago.

\bibliography{main}

\providecommand{\latin}[1]{#1}
\makeatletter
\providecommand{\doi}
  {\begingroup\let\do\@makeother\dospecials
  \catcode`\{=1 \catcode`\}=2 \doi@aux}
\providecommand{\doi@aux}[1]{\endgroup\texttt{#1}}
\makeatother
\providecommand*\mcitethebibliography{\thebibliography}
\csname @ifundefined\endcsname{endmcitethebibliography}  {\let\endmcitethebibliography\endthebibliography}{}
\begin{mcitethebibliography}{107}
\providecommand*\natexlab[1]{#1}
\providecommand*\mciteSetBstSublistMode[1]{}
\providecommand*\mciteSetBstMaxWidthForm[2]{}
\providecommand*\mciteBstWouldAddEndPuncttrue
  {\def\EndOfBibitem{\unskip.}}
\providecommand*\mciteBstWouldAddEndPunctfalse
  {\let\EndOfBibitem\relax}
\providecommand*\mciteSetBstMidEndSepPunct[3]{}
\providecommand*\mciteSetBstSublistLabelBeginEnd[3]{}
\providecommand*\EndOfBibitem{}
\mciteSetBstSublistMode{f}
\mciteSetBstMaxWidthForm{subitem}{(\alph{mcitesubitemcount})}
\mciteSetBstSublistLabelBeginEnd
  {\mcitemaxwidthsubitemform\space}
  {\relax}
  {\relax}

\bibitem[Jahnke \latin{et~al.}(2020)Jahnke, Hergenhahn, Winter, D{\"o}rner, Fr{\"u}hling, Demekhin, Gokhberg, Cederbaum, Ehresmann, Knie, and Dreuw]{doi:10.1021/acs.chemrev.0c00106}
Jahnke,~T.; Hergenhahn,~U.; Winter,~B.; D{\"o}rner,~R.; Fr{\"u}hling,~U.; Demekhin,~P.~V.; Gokhberg,~K.; Cederbaum,~L.~S.; Ehresmann,~A.; Knie,~A.; Dreuw,~A. Interatomic and Intermolecular Coulombic Decay. \emph{Chemical Reviews} \textbf{2020}, \emph{120}, 11295--11369, PMID: 33035051\relax
\mciteBstWouldAddEndPuncttrue
\mciteSetBstMidEndSepPunct{\mcitedefaultmidpunct}
{\mcitedefaultendpunct}{\mcitedefaultseppunct}\relax
\EndOfBibitem
\bibitem[Ismail \latin{et~al.}(2023)Ismail, Fert\'e, Penent, Guillemin, Peng, Marchenko, Travnikova, Inhester, Ta\"{\i}eb, Verma, Velasquez, Kukk, Trinter, Koulentianos, Mazza, Baumann, Rivas, Ovcharenko, Boll, Dold, De~Fanis, Ilchen, Meyer, Goldsztejn, Li, Doumy, Young, Sansone, D\"orner, Piancastelli, Carniato, Bozek, P\"uttner, and Simon]{PhysRevLett.131.253201}
Ismail,~I. \latin{et~al.}  Alternative Pathway to Double-Core-Hole States. \emph{Physical Review Letters} \textbf{2023}, \emph{131}, 253201\relax
\mciteBstWouldAddEndPuncttrue
\mciteSetBstMidEndSepPunct{\mcitedefaultmidpunct}
{\mcitedefaultendpunct}{\mcitedefaultseppunct}\relax
\EndOfBibitem
\bibitem[Pelimanni \latin{et~al.}(2024)Pelimanni, Fouda, Ho, Baumann, Bokarev, Fanis, Dold, Grell, Ismail, Koulentianos, Mazza, Meyer, Piancastelli, P{\"u}ttner, Rivas, Senfftleben, Simon, Young, and Doumy]{Pelimanni2024}
Pelimanni,~E. \latin{et~al.}  Observation of molecular resonant double-core excitation driven by intense X-ray pulses. \emph{Communications Physics} \textbf{2024}, \emph{7}, 341\relax
\mciteBstWouldAddEndPuncttrue
\mciteSetBstMidEndSepPunct{\mcitedefaultmidpunct}
{\mcitedefaultendpunct}{\mcitedefaultseppunct}\relax
\EndOfBibitem
\bibitem[Thompson \latin{et~al.}(2024)Thompson, Plekan, Bonanomi, Pal, Allum, Brynes, Coreno, Coriani, Danailov, Decleva, Demidovich, Devetta, Faccialà, Feifel, Forbes, Grazioli, Holland, Piseri, Prince, Rolles, Schuurman, Simoncig, Squibb, Tenorio, Vozzi, Zangrando, Callegari, Minns, and Fraia]{Thompson_2024}
Thompson,~H.~J. \latin{et~al.}  Time-resolved Auger–Meitner spectroscopy of the photodissociation dynamics of CS2. \emph{Journal of Physics B: Atomic, Molecular and Optical Physics} \textbf{2024}, \emph{57}, 215602\relax
\mciteBstWouldAddEndPuncttrue
\mciteSetBstMidEndSepPunct{\mcitedefaultmidpunct}
{\mcitedefaultendpunct}{\mcitedefaultseppunct}\relax
\EndOfBibitem
\bibitem[Kissin \latin{et~al.}(2021)Kissin, Ruberti, Kolorenč, and Averbukh]{D1CP00623A}
Kissin,~Y.; Ruberti,~M.; Kolorenč,~P.; Averbukh,~V. Attosecond pump–attosecond probe spectroscopy of Auger decay. \emph{Physical Chemistry Chemical Physics} \textbf{2021}, \emph{23}, 12376--12386\relax
\mciteBstWouldAddEndPuncttrue
\mciteSetBstMidEndSepPunct{\mcitedefaultmidpunct}
{\mcitedefaultendpunct}{\mcitedefaultseppunct}\relax
\EndOfBibitem
\bibitem[Li \latin{et~al.}(2022)Li, Driver, Rosenberger, Champenois, Duris, Al-Haddad, Averbukh, Barnard, Berrah, Bostedt, Bucksbaum, Coffee, DiMauro, Fang, Garratt, Gatton, Guo, Hartmann, Haxton, Helml, Huang, LaForge, Kamalov, Knurr, Lin, Lutman, MacArthur, Marangos, Nantel, Natan, Obaid, O’Neal, Shivaram, Schori, Walter, Wang, Wolf, Zhang, Kling, Marinelli, and Cryan]{doi:10.1126/science.abj2096}
Li,~S. \latin{et~al.}  Attosecond coherent electron motion in Auger-Meitner decay. \emph{Science} \textbf{2022}, \emph{375}, 285--290\relax
\mciteBstWouldAddEndPuncttrue
\mciteSetBstMidEndSepPunct{\mcitedefaultmidpunct}
{\mcitedefaultendpunct}{\mcitedefaultseppunct}\relax
\EndOfBibitem
\bibitem[Driver \latin{et~al.}(2024)Driver, Mountney, Wang, Ortmann, Al-Haddad, Berrah, Bostedt, Champenois, DiMauro, Duris, Garratt, Glownia, Guo, Haxton, Isele, Ivanov, Ji, Kamalov, Li, Lin, Marangos, Obaid, O'Neal, Rosenberger, Shivaram, Wang, Walter, Wolf, W{\"o}rner, Zhang, Bucksbaum, Kling, Landsman, Lucchese, Emmanouilidou, Marinelli, and Cryan]{Driver2024}
Driver,~T. \latin{et~al.}  Attosecond delays in X-ray molecular ionization. \emph{Nature} \textbf{2024}, \emph{632}, 762--767\relax
\mciteBstWouldAddEndPuncttrue
\mciteSetBstMidEndSepPunct{\mcitedefaultmidpunct}
{\mcitedefaultendpunct}{\mcitedefaultseppunct}\relax
\EndOfBibitem
\bibitem[Stern \latin{et~al.}(2009)Stern, Yacoby, Seidler, Nagle, Prange, Sorini, Rehr, and Joachimiak]{Stern:dz5149}
Stern,~E.~A.; Yacoby,~Y.; Seidler,~G.~T.; Nagle,~K.~P.; Prange,~M.~P.; Sorini,~A.~P.; Rehr,~J.~J.; Joachimiak,~A. {Reducing radiation damage in macromolecular crystals at synchrotron sources}. \emph{Acta Crystallographica Section D} \textbf{2009}, \emph{65}, 366--374\relax
\mciteBstWouldAddEndPuncttrue
\mciteSetBstMidEndSepPunct{\mcitedefaultmidpunct}
{\mcitedefaultendpunct}{\mcitedefaultseppunct}\relax
\EndOfBibitem
\bibitem[H{\'e}monnot and K{\"o}ster(2017)H{\'e}monnot, and K{\"o}ster]{doi:10.1021/acsnano.7b03447}
H{\'e}monnot,~C. Y.~J.; K{\"o}ster,~S. Imaging of biological materials and cells by X-ray scattering and diffraction. \emph{ACS Nano} \textbf{2017}, \emph{11}, 8542--8559\relax
\mciteBstWouldAddEndPuncttrue
\mciteSetBstMidEndSepPunct{\mcitedefaultmidpunct}
{\mcitedefaultendpunct}{\mcitedefaultseppunct}\relax
\EndOfBibitem
\bibitem[Carugo and Djinovi{\'c}~Carugo(2005)Carugo, and Djinovi{\'c}~Carugo]{Carugo2005-lr}
Carugo,~O.; Djinovi{\'c}~Carugo,~K. When X-rays modify the protein structure: radiation damage at work. \emph{Trends in Biochemical Sciences} \textbf{2005}, \emph{30}, 213--219\relax
\mciteBstWouldAddEndPuncttrue
\mciteSetBstMidEndSepPunct{\mcitedefaultmidpunct}
{\mcitedefaultendpunct}{\mcitedefaultseppunct}\relax
\EndOfBibitem
\bibitem[Howell(2023)]{doi:10.1080/09553002.2020.1831706}
Howell,~R.~W. Advancements in the use of Auger electrons in science and medicine during the period 2015–2019. \emph{International Journal of Radiation Biology} \textbf{2023}, \emph{99}, 2--27, PMID: 33021416\relax
\mciteBstWouldAddEndPuncttrue
\mciteSetBstMidEndSepPunct{\mcitedefaultmidpunct}
{\mcitedefaultendpunct}{\mcitedefaultseppunct}\relax
\EndOfBibitem
\bibitem[Ku \latin{et~al.}(2019)Ku, Facca, Cai, and Reilly]{Ku2019}
Ku,~A.; Facca,~V.~J.; Cai,~Z.; Reilly,~R.~M. Auger electrons for cancer therapy -- a review. \emph{EJNMMI Radiopharmacy and Chemistry} \textbf{2019}, \emph{4}, 27\relax
\mciteBstWouldAddEndPuncttrue
\mciteSetBstMidEndSepPunct{\mcitedefaultmidpunct}
{\mcitedefaultendpunct}{\mcitedefaultseppunct}\relax
\EndOfBibitem
\bibitem[Carravetta \latin{et~al.}(2000)Carravetta, Ågren, Vahtras, and Jensen]{10.1063/1.1316046}
Carravetta,~V.; Ågren,~H.; Vahtras,~O.; Jensen,~H. J.~A. {Ab initio calculations of molecular resonant photoemission spectra}. \emph{The Journal of Chemical Physics} \textbf{2000}, \emph{113}, 7790--7798\relax
\mciteBstWouldAddEndPuncttrue
\mciteSetBstMidEndSepPunct{\mcitedefaultmidpunct}
{\mcitedefaultendpunct}{\mcitedefaultseppunct}\relax
\EndOfBibitem
\bibitem[Schimmelpfennig \latin{et~al.}(1992)Schimmelpfennig, Nestmann, and Peyerimhoff]{BSchimmelpfennig_1992}
Schimmelpfennig,~B.; Nestmann,~B.; Peyerimhoff,~S.~D. Ab initio calculation of partial linewidths in the Auger decay of K-shell excited HCl. \emph{Journal of Physics B: Atomic, Molecular and Optical Physics} \textbf{1992}, \emph{25}, 1217\relax
\mciteBstWouldAddEndPuncttrue
\mciteSetBstMidEndSepPunct{\mcitedefaultmidpunct}
{\mcitedefaultendpunct}{\mcitedefaultseppunct}\relax
\EndOfBibitem
\bibitem[Schimmelpfennig \latin{et~al.}(1995)Schimmelpfennig, Nestmann, and Peyerimhoff]{SCHIMMELPFENNIG1995173}
Schimmelpfennig,~B.; Nestmann,~B.; Peyerimhoff,~S. Ab initio calculation of transition rates for autoionization: the Auger spectra of HF and F$^-$. \emph{Journal of Electron Spectroscopy and Related Phenomena} \textbf{1995}, \emph{74}, 173--186\relax
\mciteBstWouldAddEndPuncttrue
\mciteSetBstMidEndSepPunct{\mcitedefaultmidpunct}
{\mcitedefaultendpunct}{\mcitedefaultseppunct}\relax
\EndOfBibitem
\bibitem[Liegener(1982)]{LIEGENER1982188}
Liegener,~C.-M. Auger spectra by the Green's function method. \emph{Chemical Physics Letters} \textbf{1982}, \emph{90}, 188--192\relax
\mciteBstWouldAddEndPuncttrue
\mciteSetBstMidEndSepPunct{\mcitedefaultmidpunct}
{\mcitedefaultendpunct}{\mcitedefaultseppunct}\relax
\EndOfBibitem
\bibitem[Averbukh and Cederbaum(2005)Averbukh, and Cederbaum]{10.1063/1.2126976}
Averbukh,~V.; Cederbaum,~L.~S. {Ab initio calculation of interatomic decay rates by a combination of the Fano ansatz, Green’s-function methods, and the Stieltjes imaging technique}. \emph{The Journal of Chemical Physics} \textbf{2005}, \emph{123}, 204107\relax
\mciteBstWouldAddEndPuncttrue
\mciteSetBstMidEndSepPunct{\mcitedefaultmidpunct}
{\mcitedefaultendpunct}{\mcitedefaultseppunct}\relax
\EndOfBibitem
\bibitem[Skomorowski and Krylov(2021)Skomorowski, and Krylov]{10.1063/5.0036976}
Skomorowski,~W.; Krylov,~A.~I. {Feshbach–Fano approach for calculation of Auger decay rates using equation-of-motion coupled-cluster wave functions. I. Theory and implementation}. \emph{The Journal of Chemical Physics} \textbf{2021}, \emph{154}, 084124\relax
\mciteBstWouldAddEndPuncttrue
\mciteSetBstMidEndSepPunct{\mcitedefaultmidpunct}
{\mcitedefaultendpunct}{\mcitedefaultseppunct}\relax
\EndOfBibitem
\bibitem[Siegbahn \latin{et~al.}(1975)Siegbahn, Asplund, and Kelfve]{SIEGBAHN1975330}
Siegbahn,~H.; Asplund,~L.; Kelfve,~P. The Auger electron spectrum of water vapour. \emph{Chemical Physics Letters} \textbf{1975}, \emph{35}, 330--335\relax
\mciteBstWouldAddEndPuncttrue
\mciteSetBstMidEndSepPunct{\mcitedefaultmidpunct}
{\mcitedefaultendpunct}{\mcitedefaultseppunct}\relax
\EndOfBibitem
\bibitem[Jennison(1980)]{JENNISON1980435}
Jennison,~D.~R. The calculation of molecular and cluster auger spectra. \emph{Chemical Physics Letters} \textbf{1980}, \emph{69}, 435--440\relax
\mciteBstWouldAddEndPuncttrue
\mciteSetBstMidEndSepPunct{\mcitedefaultmidpunct}
{\mcitedefaultendpunct}{\mcitedefaultseppunct}\relax
\EndOfBibitem
\bibitem[Larkins \latin{et~al.}(1990)Larkins, Tulea, and Chelkowska]{Larkins1990}
Larkins,~F.~P.; Tulea,~L.~C.; Chelkowska,~E.~Z. Auger Electron Spectra of Molecules: The First Row Hydrides. \emph{Australian Journal of Physics} \textbf{1990}, \emph{43}, 625--640\relax
\mciteBstWouldAddEndPuncttrue
\mciteSetBstMidEndSepPunct{\mcitedefaultmidpunct}
{\mcitedefaultendpunct}{\mcitedefaultseppunct}\relax
\EndOfBibitem
\bibitem[Fink(1995)]{FINK1995295}
Fink,~R. Theoretical autoionization spectra of 1s \(\to\) \(\pi^\ast\) excited N2 and N2O. \emph{Journal of Electron Spectroscopy and Related Phenomena} \textbf{1995}, \emph{76}, 295--300, Proceedings of the Sixth International Conference on Electron Spectroscopy\relax
\mciteBstWouldAddEndPuncttrue
\mciteSetBstMidEndSepPunct{\mcitedefaultmidpunct}
{\mcitedefaultendpunct}{\mcitedefaultseppunct}\relax
\EndOfBibitem
\bibitem[Travnikova \latin{et~al.}(2009)Travnikova, Fink, Kivimäki, Céolin, Bao, and Piancastelli]{TRAVNIKOVA200967}
Travnikova,~O.; Fink,~R.~F.; Kivimäki,~A.; Céolin,~D.; Bao,~Z.; Piancastelli,~M.~N. Assignment of the L2,3VV normal Auger decay spectrum of Cl2 by ab initio calculations. \emph{Chemical Physics Letters} \textbf{2009}, \emph{474}, 67--73\relax
\mciteBstWouldAddEndPuncttrue
\mciteSetBstMidEndSepPunct{\mcitedefaultmidpunct}
{\mcitedefaultendpunct}{\mcitedefaultseppunct}\relax
\EndOfBibitem
\bibitem[Inhester \latin{et~al.}(2016)Inhester, Hanasaki, Hao, Son, and Santra]{PhysRevA.94.023422}
Inhester,~L.; Hanasaki,~K.; Hao,~Y.; Son,~S.-K.; Santra,~R. X-ray multiphoton ionization dynamics of a water molecule irradiated by an x-ray free-electron laser pulse. \emph{Physical Review A} \textbf{2016}, \emph{94}, 023422\relax
\mciteBstWouldAddEndPuncttrue
\mciteSetBstMidEndSepPunct{\mcitedefaultmidpunct}
{\mcitedefaultendpunct}{\mcitedefaultseppunct}\relax
\EndOfBibitem
\bibitem[Hao \latin{et~al.}(2015)Hao, Inhester, Hanasaki, Son, and Santra]{10.1063/1.4919794}
Hao,~Y.; Inhester,~L.; Hanasaki,~K.; Son,~S.-K.; Santra,~R. {Efficient electronic structure calculation for molecular ionization dynamics at high x-ray intensity}. \emph{Structural Dynamics} \textbf{2015}, \emph{2}, 041707\relax
\mciteBstWouldAddEndPuncttrue
\mciteSetBstMidEndSepPunct{\mcitedefaultmidpunct}
{\mcitedefaultendpunct}{\mcitedefaultseppunct}\relax
\EndOfBibitem
\bibitem[Faegri and Kelly(1979)Faegri, and Kelly]{PhysRevA.19.1649}
Faegri,~K.; Kelly,~H.~P. Calculated Auger transition rates for HF. \emph{Physical Review A} \textbf{1979}, \emph{19}, 1649--1655\relax
\mciteBstWouldAddEndPuncttrue
\mciteSetBstMidEndSepPunct{\mcitedefaultmidpunct}
{\mcitedefaultendpunct}{\mcitedefaultseppunct}\relax
\EndOfBibitem
\bibitem[Higashi \latin{et~al.}(1982)Higashi, Hiroike, and Nakajima]{HIGASHI1982377}
Higashi,~M.; Hiroike,~E.; Nakajima,~T. Calculations of the Auger transition rates in molecules. I. Effect of the nonspherical potential: Application to CH4. \emph{Chemical Physics} \textbf{1982}, \emph{68}, 377--382\relax
\mciteBstWouldAddEndPuncttrue
\mciteSetBstMidEndSepPunct{\mcitedefaultmidpunct}
{\mcitedefaultendpunct}{\mcitedefaultseppunct}\relax
\EndOfBibitem
\bibitem[Z\"ahringer \latin{et~al.}(1992)Z\"ahringer, Meyer, and Cederbaum]{PhysRevA.45.318}
Z\"ahringer,~K.; Meyer,~H.-D.; Cederbaum,~L.~S. Molecular scattering wave functions for Auger decay rates: The Auger spectrum of hydrogen fluoride. \emph{Physical Review A} \textbf{1992}, \emph{45}, 318--328\relax
\mciteBstWouldAddEndPuncttrue
\mciteSetBstMidEndSepPunct{\mcitedefaultmidpunct}
{\mcitedefaultendpunct}{\mcitedefaultseppunct}\relax
\EndOfBibitem
\bibitem[Demekhin \latin{et~al.}(2007)Demekhin, Omel'yanenko, Lagutin, Sukhorukov, Werner, Ehresmann, Schartner, and Schmoranzer]{Demekhin2007}
Demekhin,~P.~V.; Omel'yanenko,~D.~V.; Lagutin,~B.~M.; Sukhorukov,~V.~L.; Werner,~L.; Ehresmann,~A.; Schartner,~K.-H.; Schmoranzer,~H. Investigation of photoionization and photodissociation of an oxygen molecule by the method of coupled differential equations. \emph{Optics and Spectroscopy} \textbf{2007}, \emph{102}, 318--329\relax
\mciteBstWouldAddEndPuncttrue
\mciteSetBstMidEndSepPunct{\mcitedefaultmidpunct}
{\mcitedefaultendpunct}{\mcitedefaultseppunct}\relax
\EndOfBibitem
\bibitem[Demekhin \latin{et~al.}(2009)Demekhin, Petrov, Sukhorukov, Kielich, Reiss, Hentges, Haar, Schmoranzer, and Ehresmann]{PhysRevA.80.063425}
Demekhin,~P.~V.; Petrov,~I.~D.; Sukhorukov,~V.~L.; Kielich,~W.; Reiss,~P.; Hentges,~R.; Haar,~I.; Schmoranzer,~H.; Ehresmann,~A. Interference effects during the Auger decay of the ${\text{C}}^{\ensuremath{\ast}}\text{O}$ $1{s}^{\ensuremath{-}1}{\ensuremath{\pi}}^{\ensuremath{\ast}}$ resonance studied by angular distribution of the ${\text{CO}}^{+}(A)$ photoelectrons and polarization analysis of the ${\text{CO}}^{+}(A\text{\ensuremath{-}}X)$ fluorescence. \emph{Physical Review A} \textbf{2009}, \emph{80}, 063425\relax
\mciteBstWouldAddEndPuncttrue
\mciteSetBstMidEndSepPunct{\mcitedefaultmidpunct}
{\mcitedefaultendpunct}{\mcitedefaultseppunct}\relax
\EndOfBibitem
\bibitem[Demekhin \latin{et~al.}(2011)Demekhin, Ehresmann, and Sukhorukov]{10.1063/1.3526026}
Demekhin,~P.~V.; Ehresmann,~A.; Sukhorukov,~V.~L. {Single center method: A computational tool for ionization and electronic excitation studies of molecules}. \emph{The Journal of Chemical Physics} \textbf{2011}, \emph{134}, 024113\relax
\mciteBstWouldAddEndPuncttrue
\mciteSetBstMidEndSepPunct{\mcitedefaultmidpunct}
{\mcitedefaultendpunct}{\mcitedefaultseppunct}\relax
\EndOfBibitem
\bibitem[Inhester \latin{et~al.}(2012)Inhester, Burmeister, Groenhof, and Grubmüller]{10.1063/1.3700233}
Inhester,~L.; Burmeister,~C.~F.; Groenhof,~G.; Grubmüller,~H. {Auger spectrum of a water molecule after single and double core ionization}. \emph{The Journal of Chemical Physics} \textbf{2012}, \emph{136}, 144304\relax
\mciteBstWouldAddEndPuncttrue
\mciteSetBstMidEndSepPunct{\mcitedefaultmidpunct}
{\mcitedefaultendpunct}{\mcitedefaultseppunct}\relax
\EndOfBibitem
\bibitem[Banks \latin{et~al.}(2017)Banks, Little, Tennyson, and Emmanouilidou]{C7CP02345F}
Banks,~H. I.~B.; Little,~D.~A.; Tennyson,~J.; Emmanouilidou,~A. Interaction of molecular nitrogen with free-electron-laser radiation. \emph{Physical Chemistry Chemical Physics} \textbf{2017}, \emph{19}, 19794--19806\relax
\mciteBstWouldAddEndPuncttrue
\mciteSetBstMidEndSepPunct{\mcitedefaultmidpunct}
{\mcitedefaultendpunct}{\mcitedefaultseppunct}\relax
\EndOfBibitem
\bibitem[Mitani \latin{et~al.}(2003)Mitani, Takahashi, Saito, and Iwata]{MITANI2003103}
Mitani,~M.; Takahashi,~O.; Saito,~K.; Iwata,~S. Theoretical molecular Auger spectra with electron population analysis. \emph{Journal of Electron Spectroscopy and Related Phenomena} \textbf{2003}, \emph{128}, 103--117\relax
\mciteBstWouldAddEndPuncttrue
\mciteSetBstMidEndSepPunct{\mcitedefaultmidpunct}
{\mcitedefaultendpunct}{\mcitedefaultseppunct}\relax
\EndOfBibitem
\bibitem[Takahashi \latin{et~al.}(2006)Takahashi, Odelius, Nordlund, Nilsson, Bluhm, and Pettersson]{10.1063/1.2166234}
Takahashi,~O.; Odelius,~M.; Nordlund,~D.; Nilsson,~A.; Bluhm,~H.; Pettersson,~L. G.~M. {Auger decay calculations with core-hole excited-state molecular-dynamics simulations of water}. \emph{The Journal of Chemical Physics} \textbf{2006}, \emph{124}, 064307\relax
\mciteBstWouldAddEndPuncttrue
\mciteSetBstMidEndSepPunct{\mcitedefaultmidpunct}
{\mcitedefaultendpunct}{\mcitedefaultseppunct}\relax
\EndOfBibitem
\bibitem[de~Moura \latin{et~al.}(2023)de~Moura, Laurent, Bozek, Briant, Çarçabal, Cubaynes, Shafizadeh, Simon, Soep, Püttner, and et~al.]{D3CP01746J}
de~Moura,~C. E.~V.; Laurent,~J.; Bozek,~J.; Briant,~M.; Çarçabal,~P.; Cubaynes,~D.; Shafizadeh,~N.; Simon,~M.; Soep,~B.; Püttner,~R.; et~al. Experimental and theoretical study of resonant core-hole spectroscopies of gas-phase free-base phthalocyanine. \emph{Physical Chemistry Chemical Physics} \textbf{2023}, \emph{25}, 15555--15566\relax
\mciteBstWouldAddEndPuncttrue
\mciteSetBstMidEndSepPunct{\mcitedefaultmidpunct}
{\mcitedefaultendpunct}{\mcitedefaultseppunct}\relax
\EndOfBibitem
\bibitem[Fouda \latin{et~al.}(2023)Fouda, Koulentianos, Young, Doumy, and Ho]{doi:10.1080/00268976.2022.2133749}
Fouda,~A. E.~A.; Koulentianos,~D.; Young,~L.; Doumy,~G.; Ho,~P.~J. Resonant double-core excitations with ultrafast, intense X-ray pulses. \emph{Molecular Physics} \textbf{2023}, \emph{121}, e2133749\relax
\mciteBstWouldAddEndPuncttrue
\mciteSetBstMidEndSepPunct{\mcitedefaultmidpunct}
{\mcitedefaultendpunct}{\mcitedefaultseppunct}\relax
\EndOfBibitem
\bibitem[Fouda \latin{et~al.}(2024)Fouda, Lindblom, Southworth, Doumy, Ho, Young, Cheng, and Sorensen]{doi:10.1021/acs.jpclett.3c03611}
Fouda,~A. E.~A.; Lindblom,~V.; Southworth,~S.~H.; Doumy,~G.; Ho,~P.~J.; Young,~L.; Cheng,~L.; Sorensen,~S.~L. Influence of Selective Carbon 1s Excitation on Auger–Meitner Decay in the ESCA Molecule. \emph{The Journal of Physical Chemistry Letters} \textbf{2024}, \emph{15}, 4286--4293, PMID: 38608168\relax
\mciteBstWouldAddEndPuncttrue
\mciteSetBstMidEndSepPunct{\mcitedefaultmidpunct}
{\mcitedefaultendpunct}{\mcitedefaultseppunct}\relax
\EndOfBibitem
\bibitem[Deleuze \latin{et~al.}(2001)Deleuze, Trofimov, and Cederbaum]{doi:10.1063/1.1386414}
Deleuze,~M.~S.; Trofimov,~A.~B.; Cederbaum,~L.~S. Valence one-electron and shake-up ionization bands of polycyclic aromatic hydrocarbons. I. Benzene, naphthalene, anthracene, naphthacene, and pentacene. \emph{The Journal of Chemical Physics} \textbf{2001}, \emph{115}, 5859--5882\relax
\mciteBstWouldAddEndPuncttrue
\mciteSetBstMidEndSepPunct{\mcitedefaultmidpunct}
{\mcitedefaultendpunct}{\mcitedefaultseppunct}\relax
\EndOfBibitem
\bibitem[Werner and Meyer(1981)Werner, and Meyer]{werner1981quadratically}
Werner,~H.-J.; Meyer,~W. A quadratically convergent MCSCF method for the simultaneous optimization of several states. \emph{The Journal of Chemical Physics} \textbf{1981}, \emph{74}, 5794--5801\relax
\mciteBstWouldAddEndPuncttrue
\mciteSetBstMidEndSepPunct{\mcitedefaultmidpunct}
{\mcitedefaultendpunct}{\mcitedefaultseppunct}\relax
\EndOfBibitem
\bibitem[Malmqvist \latin{et~al.}(1990)Malmqvist, Rendell, and Roos]{malmqvist1990restricted}
Malmqvist,~P.~{\AA}.; Rendell,~A.; Roos,~B.~O. The restricted active space self-consistent-field method, implemented with a split graph unitary group approach. \emph{Journal of Physical Chemistry} \textbf{1990}, \emph{94}, 5477--5482\relax
\mciteBstWouldAddEndPuncttrue
\mciteSetBstMidEndSepPunct{\mcitedefaultmidpunct}
{\mcitedefaultendpunct}{\mcitedefaultseppunct}\relax
\EndOfBibitem
\bibitem[Bokarev and Kühn(2020)Bokarev, and Kühn]{https://doi.org/10.1002/wcms.1433}
Bokarev,~S.~I.; Kühn,~O. Theoretical X-ray spectroscopy of transition metal compounds. \emph{WIREs Computational Molecular Science} \textbf{2020}, \emph{10}, e1433\relax
\mciteBstWouldAddEndPuncttrue
\mciteSetBstMidEndSepPunct{\mcitedefaultmidpunct}
{\mcitedefaultendpunct}{\mcitedefaultseppunct}\relax
\EndOfBibitem
\bibitem[Pinjari \latin{et~al.}(2016)Pinjari, Delcey, Guo, Odelius, and Lundberg]{https://doi.org/10.1002/jcc.24237}
Pinjari,~R.~V.; Delcey,~M.~G.; Guo,~M.; Odelius,~M.; Lundberg,~M. Cost and sensitivity of restricted active-space calculations of metal L-edge X-ray absorption spectra. \emph{Journal of Computational Chemistry} \textbf{2016}, \emph{37}, 477--486\relax
\mciteBstWouldAddEndPuncttrue
\mciteSetBstMidEndSepPunct{\mcitedefaultmidpunct}
{\mcitedefaultendpunct}{\mcitedefaultseppunct}\relax
\EndOfBibitem
\bibitem[Josefsson \latin{et~al.}(2012)Josefsson, Kunnus, Schreck, F{\"o}hlisch, de~Groot, Wernet, and Odelius]{doi:10.1021/jz301479j}
Josefsson,~I.; Kunnus,~K.; Schreck,~S.; F{\"o}hlisch,~A.; de~Groot,~F.; Wernet,~P.; Odelius,~M. Ab Initio Calculations of X-ray Spectra: Atomic Multiplet and Molecular Orbital Effects in a Multiconfigurational SCF Approach to the L-Edge Spectra of Transition Metal Complexes. \emph{The Journal of Physical Chemistry Letters} \textbf{2012}, \emph{3}, 3565--3570, PMID: 26290989\relax
\mciteBstWouldAddEndPuncttrue
\mciteSetBstMidEndSepPunct{\mcitedefaultmidpunct}
{\mcitedefaultendpunct}{\mcitedefaultseppunct}\relax
\EndOfBibitem
\bibitem[Grell \latin{et~al.}(2015)Grell, Bokarev, Winter, Seidel, Aziz, Aziz, and Kühn]{10.1063/1.4928511}
Grell,~G.; Bokarev,~S.~I.; Winter,~B.; Seidel,~R.; Aziz,~E.~F.; Aziz,~S.~G.; Kühn,~O. {Multi-reference approach to the calculation of photoelectron spectra including spin-orbit coupling}. \emph{The Journal of Chemical Physics} \textbf{2015}, \emph{143}, 074104\relax
\mciteBstWouldAddEndPuncttrue
\mciteSetBstMidEndSepPunct{\mcitedefaultmidpunct}
{\mcitedefaultendpunct}{\mcitedefaultseppunct}\relax
\EndOfBibitem
\bibitem[Northey \latin{et~al.}(2020)Northey, Norell, Fouda, Besley, Odelius, and Penfold]{C9CP03019K}
Northey,~T.; Norell,~J.; Fouda,~A. E.~A.; Besley,~N.~A.; Odelius,~M.; Penfold,~T.~J. Ultrafast nonadiabatic dynamics probed by nitrogen K-edge absorption spectroscopy. \emph{Physical Chemistry Chemical Physics} \textbf{2020}, \emph{22}, 2667--2676\relax
\mciteBstWouldAddEndPuncttrue
\mciteSetBstMidEndSepPunct{\mcitedefaultmidpunct}
{\mcitedefaultendpunct}{\mcitedefaultseppunct}\relax
\EndOfBibitem
\bibitem[Fouda \latin{et~al.}(2020)Fouda, Seitz, Hauschild, Blum, Yang, Heske, Weinhardt, and Besley]{doi:10.1021/acs.jpclett.0c01981}
Fouda,~A. E.~A.; Seitz,~L.~C.; Hauschild,~D.; Blum,~M.; Yang,~W.; Heske,~C.; Weinhardt,~L.; Besley,~N.~A. Observation of Double Excitations in the Resonant Inelastic X-ray Scattering of Nitric Oxide. \emph{The Journal of Physical Chemistry Letters} \textbf{2020}, \emph{11}, 7476--7482, PMID: 32787301\relax
\mciteBstWouldAddEndPuncttrue
\mciteSetBstMidEndSepPunct{\mcitedefaultmidpunct}
{\mcitedefaultendpunct}{\mcitedefaultseppunct}\relax
\EndOfBibitem
\bibitem[Koulentianos \latin{et~al.}(2020)Koulentianos, Fouda, Southworth, Bozek, Küpper, Santra, Kryzhevoi, Cederbaum, Bostedt, Messerschmidt, Berrah, Fang, Murphy, Osipov, Cryan, Glownia, Ghimire, Ho, Krässig, Ray, Li, Kanter, Young, and Doumy]{Koulentianos_2020}
Koulentianos,~D. \latin{et~al.}  High intensity x-ray interaction with a model bio-molecule system: double-core-hole states and fragmentation of formamide. \emph{Journal of Physics B: Atomic, Molecular and Optical Physics} \textbf{2020}, \emph{53}, 244005\relax
\mciteBstWouldAddEndPuncttrue
\mciteSetBstMidEndSepPunct{\mcitedefaultmidpunct}
{\mcitedefaultendpunct}{\mcitedefaultseppunct}\relax
\EndOfBibitem
\bibitem[Fouda and Ho(2021)Fouda, and Ho]{10.1063/5.0050891}
Fouda,~A. E.~A.; Ho,~P.~J. {Site-specific generation of excited state wavepackets with high-intensity attosecond x rays}. \emph{The Journal of Chemical Physics} \textbf{2021}, \emph{154}, 224111\relax
\mciteBstWouldAddEndPuncttrue
\mciteSetBstMidEndSepPunct{\mcitedefaultmidpunct}
{\mcitedefaultendpunct}{\mcitedefaultseppunct}\relax
\EndOfBibitem
\bibitem[Grell \latin{et~al.}(2019)Grell, K\"uhn, and Bokarev]{PhysRevA.100.042512}
Grell,~G.; K\"uhn,~O.; Bokarev,~S.~I. Multireference quantum chemistry protocol for simulating autoionization spectra: Test of ionization continuum models for the neon atom. \emph{Physical Review A} \textbf{2019}, \emph{100}, 042512\relax
\mciteBstWouldAddEndPuncttrue
\mciteSetBstMidEndSepPunct{\mcitedefaultmidpunct}
{\mcitedefaultendpunct}{\mcitedefaultseppunct}\relax
\EndOfBibitem
\bibitem[Grell and Bokarev(2020)Grell, and Bokarev]{10.1063/1.5142251}
Grell,~G.; Bokarev,~S.~I. {Multi-reference protocol for (auto)ionization spectra: Application to molecules}. \emph{The Journal of Chemical Physics} \textbf{2020}, \emph{152}, 074108\relax
\mciteBstWouldAddEndPuncttrue
\mciteSetBstMidEndSepPunct{\mcitedefaultmidpunct}
{\mcitedefaultendpunct}{\mcitedefaultseppunct}\relax
\EndOfBibitem
\bibitem[Tenorio \latin{et~al.}(2022)Tenorio, Vo{\ss}, Bokarev, Decleva, and Coriani]{cabraljctc2022}
Tenorio,~B. N.~C.; Vo{\ss},~T.~A.; Bokarev,~S.~I.; Decleva,~P.; Coriani,~S. Multireference Approach to Normal and Resonant Auger Spectra Based on the One-Center Approximation. \emph{Journal of Chemical Theory and Computation} \textbf{2022}, \emph{18}, 4387--4407, PMID: 35737643\relax
\mciteBstWouldAddEndPuncttrue
\mciteSetBstMidEndSepPunct{\mcitedefaultmidpunct}
{\mcitedefaultendpunct}{\mcitedefaultseppunct}\relax
\EndOfBibitem
\bibitem[Malmqvist \latin{et~al.}(2008)Malmqvist, Pierloot, Shahi, Cramer, and Gagliardi]{malmqvist2008restricted}
Malmqvist,~P.~{\AA}.; Pierloot,~K.; Shahi,~A. R.~M.; Cramer,~C.~J.; Gagliardi,~L. The restricted active space followed by second-order perturbation theory method: Theory and application to the study of Cu O 2 and Cu 2 O 2 systems. \emph{The Journal of Chemical Physics} \textbf{2008}, \emph{128}, 204109\relax
\mciteBstWouldAddEndPuncttrue
\mciteSetBstMidEndSepPunct{\mcitedefaultmidpunct}
{\mcitedefaultendpunct}{\mcitedefaultseppunct}\relax
\EndOfBibitem
\bibitem[Li~Manni \latin{et~al.}(2014)Li~Manni, Carlson, Luo, Ma, Olsen, Truhlar, and Gagliardi]{mcpdft2014}
Li~Manni,~G.; Carlson,~R.~K.; Luo,~S.; Ma,~D.; Olsen,~J.; Truhlar,~D.~G.; Gagliardi,~L. Multiconfiguration Pair-Density Functional Theory. \emph{Journal of Chemical Theory and Computation} \textbf{2014}, \emph{10}, 3669--3680, PMID: 26588512\relax
\mciteBstWouldAddEndPuncttrue
\mciteSetBstMidEndSepPunct{\mcitedefaultmidpunct}
{\mcitedefaultendpunct}{\mcitedefaultseppunct}\relax
\EndOfBibitem
\bibitem[Gagliardi \latin{et~al.}(2017)Gagliardi, Truhlar, Li~Manni, Carlson, Hoyer, and Bao]{mcpdftreview01}
Gagliardi,~L.; Truhlar,~D.~G.; Li~Manni,~G.; Carlson,~R.~K.; Hoyer,~C.~E.; Bao,~J.~L. Multiconfiguration pair-density functional theory: A new way to treat strongly correlated systems. \emph{Accounts of Chemical Research} \textbf{2017}, \emph{50}, 66--73\relax
\mciteBstWouldAddEndPuncttrue
\mciteSetBstMidEndSepPunct{\mcitedefaultmidpunct}
{\mcitedefaultendpunct}{\mcitedefaultseppunct}\relax
\EndOfBibitem
\bibitem[Sharma \latin{et~al.}(2021)Sharma, Bao, Truhlar, and Gagliardi]{mcpdftreview02}
Sharma,~P.; Bao,~J.~J.; Truhlar,~D.~G.; Gagliardi,~L. Multiconfiguration pair-density functional theory. \emph{Annual Review of Physical Chemistry} \textbf{2021}, \emph{72}, 541--564\relax
\mciteBstWouldAddEndPuncttrue
\mciteSetBstMidEndSepPunct{\mcitedefaultmidpunct}
{\mcitedefaultendpunct}{\mcitedefaultseppunct}\relax
\EndOfBibitem
\bibitem[Zhou \latin{et~al.}(2022)Zhou, Hermes, Wu, Bao, Pandharkar, King, Zhang, Scott, Lykhin, Gagliardi, \latin{et~al.} others]{mcpdftreview03}
Zhou,~C.; Hermes,~M.~R.; Wu,~D.; Bao,~J.~J.; Pandharkar,~R.; King,~D.~S.; Zhang,~D.; Scott,~T.~R.; Lykhin,~A.~O.; Gagliardi,~L.; others Electronic structure of strongly correlated systems: recent developments in multiconfiguration pair-density functional theory and multiconfiguration nonclassical-energy functional theory. \emph{Chemical Science} \textbf{2022}, \emph{13}, 7685--7706\relax
\mciteBstWouldAddEndPuncttrue
\mciteSetBstMidEndSepPunct{\mcitedefaultmidpunct}
{\mcitedefaultendpunct}{\mcitedefaultseppunct}\relax
\EndOfBibitem
\bibitem[Ghosh \latin{et~al.}(2023)Ghosh, Mukamel, and Govind]{ghosh2023combined}
Ghosh,~S.; Mukamel,~S.; Govind,~N. A combined wave function and density functional approach for K-edge X-ray absorption near-edge spectroscopy: A case study of hydrated first-row transition metal ions. \emph{The Journal of Physical Chemistry Letters} \textbf{2023}, \emph{14}, 5203--5209\relax
\mciteBstWouldAddEndPuncttrue
\mciteSetBstMidEndSepPunct{\mcitedefaultmidpunct}
{\mcitedefaultendpunct}{\mcitedefaultseppunct}\relax
\EndOfBibitem
\bibitem[Ghosh \latin{et~al.}(2018)Ghosh, Verma, Cramer, Gagliardi, and Truhlar]{doi:10.1021/acs.chemrev.8b00193}
Ghosh,~S.; Verma,~P.; Cramer,~C.~J.; Gagliardi,~L.; Truhlar,~D.~G. Combining Wave Function Methods with Density Functional Theory for Excited States. \emph{Chemical Reviews} \textbf{2018}, \emph{118}, 7249--7292, PMID: 30044618\relax
\mciteBstWouldAddEndPuncttrue
\mciteSetBstMidEndSepPunct{\mcitedefaultmidpunct}
{\mcitedefaultendpunct}{\mcitedefaultseppunct}\relax
\EndOfBibitem
\bibitem[Tishchenko \latin{et~al.}(2008)Tishchenko, Zheng, and Truhlar]{doi:10.1021/ct800077r}
Tishchenko,~O.; Zheng,~J.; Truhlar,~D.~G. Multireference Model Chemistries for Thermochemical Kinetics. \emph{Journal of Chemical Theory and Computation} \textbf{2008}, \emph{4}, 1208--1219, PMID: 26631697\relax
\mciteBstWouldAddEndPuncttrue
\mciteSetBstMidEndSepPunct{\mcitedefaultmidpunct}
{\mcitedefaultendpunct}{\mcitedefaultseppunct}\relax
\EndOfBibitem
\bibitem[Olsen \latin{et~al.}(1988)Olsen, Roos, Jo/rgensen, and Jensen]{olsen1988determinant}
Olsen,~J.; Roos,~B.~O.; Jo/rgensen,~P.; Jensen,~H. J.~A. Determinant based configuration interaction algorithms for complete and restricted configuration interaction spaces. \emph{The Journal of Chemical Physics} \textbf{1988}, \emph{89}, 2185--2192\relax
\mciteBstWouldAddEndPuncttrue
\mciteSetBstMidEndSepPunct{\mcitedefaultmidpunct}
{\mcitedefaultendpunct}{\mcitedefaultseppunct}\relax
\EndOfBibitem
\bibitem[Sauri \latin{et~al.}(2011)Sauri, Serrano-Andr{\'e}s, Shahi, Gagliardi, Vancoillie, and Pierloot]{sauri2011multiconfigurational}
Sauri,~V.; Serrano-Andr{\'e}s,~L.; Shahi,~A. R.~M.; Gagliardi,~L.; Vancoillie,~S.; Pierloot,~K. Multiconfigurational second-order perturbation theory restricted active space (RASPT2) method for electronic excited states: a benchmark study. \emph{Journal of Chemical Theory and Computation} \textbf{2011}, \emph{7}, 153--168\relax
\mciteBstWouldAddEndPuncttrue
\mciteSetBstMidEndSepPunct{\mcitedefaultmidpunct}
{\mcitedefaultendpunct}{\mcitedefaultseppunct}\relax
\EndOfBibitem
\bibitem[Casanova(2022)]{casanova2022restricted}
Casanova,~D. Restricted active space configuration interaction methods for strong correlation: Recent developments. \emph{Wiley Interdisciplinary Reviews: Computational Molecular Science} \textbf{2022}, \emph{12}, e1561\relax
\mciteBstWouldAddEndPuncttrue
\mciteSetBstMidEndSepPunct{\mcitedefaultmidpunct}
{\mcitedefaultendpunct}{\mcitedefaultseppunct}\relax
\EndOfBibitem
\bibitem[Delcey \latin{et~al.}(2019)Delcey, S{\o}rensen, Vacher, Couto, and Lundberg]{delcey2019efficient}
Delcey,~M.~G.; S{\o}rensen,~L.~K.; Vacher,~M.; Couto,~R.~C.; Lundberg,~M. Efficient calculations of a large number of highly excited states for multiconfigurational wavefunctions. \emph{Journal of Computational Chemistry} \textbf{2019}, \emph{40}, 1789--1799\relax
\mciteBstWouldAddEndPuncttrue
\mciteSetBstMidEndSepPunct{\mcitedefaultmidpunct}
{\mcitedefaultendpunct}{\mcitedefaultseppunct}\relax
\EndOfBibitem
\bibitem[Fdez.~Galván \latin{et~al.}(2019)Fdez.~Galván, Vacher, Alavi, Angeli, Aquilante, Autschbach, Bao, Bokarev, Bogdanov, Carlson, and et~al.]{fdez2019openmolcas}
Fdez.~Galván,~I.; Vacher,~M.; Alavi,~A.; Angeli,~C.; Aquilante,~F.; Autschbach,~J.; Bao,~J.~J.; Bokarev,~S.~I.; Bogdanov,~N.~A.; Carlson,~R.~K.; et~al. OpenMolcas: From Source Code to Insight. \emph{Journal of Chemical Theory and Computation} \textbf{2019}, \emph{15}, 5925--5964, PMID: 31509407\relax
\mciteBstWouldAddEndPuncttrue
\mciteSetBstMidEndSepPunct{\mcitedefaultmidpunct}
{\mcitedefaultendpunct}{\mcitedefaultseppunct}\relax
\EndOfBibitem
\bibitem[Pulay(2011)]{pulay2011perspective}
Pulay,~P. A perspective on the CASPT2 method. \emph{International Journal of Quantum Chemistry} \textbf{2011}, \emph{111}, 3273--3279\relax
\mciteBstWouldAddEndPuncttrue
\mciteSetBstMidEndSepPunct{\mcitedefaultmidpunct}
{\mcitedefaultendpunct}{\mcitedefaultseppunct}\relax
\EndOfBibitem
\bibitem[Roos and Andersson(1995)Roos, and Andersson]{roos1995multiconfigurational}
Roos,~B.~O.; Andersson,~K. Multiconfigurational perturbation theory with level shift—the Cr2 potential revisited. \emph{Chemical Physics Letters} \textbf{1995}, \emph{245}, 215--223\relax
\mciteBstWouldAddEndPuncttrue
\mciteSetBstMidEndSepPunct{\mcitedefaultmidpunct}
{\mcitedefaultendpunct}{\mcitedefaultseppunct}\relax
\EndOfBibitem
\bibitem[Forsberg and Malmqvist(1997)Forsberg, and Malmqvist]{forsberg1997multiconfiguration}
Forsberg,~N.; Malmqvist,~P.-{\AA}. Multiconfiguration perturbation theory with imaginary level shift. \emph{Chemical Physics Letters} \textbf{1997}, \emph{274}, 196--204\relax
\mciteBstWouldAddEndPuncttrue
\mciteSetBstMidEndSepPunct{\mcitedefaultmidpunct}
{\mcitedefaultendpunct}{\mcitedefaultseppunct}\relax
\EndOfBibitem
\bibitem[Ghigo \latin{et~al.}(2004)Ghigo, Roos, and Malmqvist]{ghigo2004modified}
Ghigo,~G.; Roos,~B.~O.; Malmqvist,~P.-{\AA}. A modified definition of the zeroth-order Hamiltonian in multiconfigurational perturbation theory (CASPT2). \emph{Chemical Physics Letters} \textbf{2004}, \emph{396}, 142--149\relax
\mciteBstWouldAddEndPuncttrue
\mciteSetBstMidEndSepPunct{\mcitedefaultmidpunct}
{\mcitedefaultendpunct}{\mcitedefaultseppunct}\relax
\EndOfBibitem
\bibitem[Zobel \latin{et~al.}(2017)Zobel, Nogueira, and Gonz{\'a}lez]{zobel2017ipea}
Zobel,~J.~P.; Nogueira,~J.~J.; Gonz{\'a}lez,~L. The IPEA dilemma in CASPT2. \emph{Chemical Science} \textbf{2017}, \emph{8}, 1482--1499\relax
\mciteBstWouldAddEndPuncttrue
\mciteSetBstMidEndSepPunct{\mcitedefaultmidpunct}
{\mcitedefaultendpunct}{\mcitedefaultseppunct}\relax
\EndOfBibitem
\bibitem[Guo \latin{et~al.}(2016)Guo, Sivalingam, Valeev, and Neese]{guo2016sparsemaps}
Guo,~Y.; Sivalingam,~K.; Valeev,~E.~F.; Neese,~F. SparseMaps—A systematic infrastructure for reduced-scaling electronic structure methods. III. Linear-scaling multireference domain-based pair natural orbital N-electron valence perturbation theory. \emph{The Journal of Chemical Physics} \textbf{2016}, \emph{144}\relax
\mciteBstWouldAddEndPuncttrue
\mciteSetBstMidEndSepPunct{\mcitedefaultmidpunct}
{\mcitedefaultendpunct}{\mcitedefaultseppunct}\relax
\EndOfBibitem
\bibitem[Manna \latin{et~al.}(2024)Manna, Jangid, Pant, and Dutta]{jangid2024efficient}
Manna,~A.; Jangid,~B.; Pant,~R.; Dutta,~A.~K. Efficient State-Specific Natural Orbital Based Equation of Motion Coupled Cluster Method for Core-Ionization Energies: Theory, Implementation, and Benchmark. \emph{Journal of Chemical Theory and Computation} \textbf{2024}, \emph{20}, 6604--6620\relax
\mciteBstWouldAddEndPuncttrue
\mciteSetBstMidEndSepPunct{\mcitedefaultmidpunct}
{\mcitedefaultendpunct}{\mcitedefaultseppunct}\relax
\EndOfBibitem
\bibitem[Carlson \latin{et~al.}(2015)Carlson, Li~Manni, Sonnenberger, Truhlar, and Gagliardi]{carlson2015multiconfiguration}
Carlson,~R.~K.; Li~Manni,~G.; Sonnenberger,~A.~L.; Truhlar,~D.~G.; Gagliardi,~L. Multiconfiguration pair-density functional theory: Barrier heights and main group and transition metal energetics. \emph{Journal of Chemical Theory and Computation} \textbf{2015}, \emph{11}, 82--90\relax
\mciteBstWouldAddEndPuncttrue
\mciteSetBstMidEndSepPunct{\mcitedefaultmidpunct}
{\mcitedefaultendpunct}{\mcitedefaultseppunct}\relax
\EndOfBibitem
\bibitem[Pandharkar \latin{et~al.}(2020)Pandharkar, Hermes, Truhlar, and Gagliardi]{pandharkar2020new}
Pandharkar,~R.; Hermes,~M.~R.; Truhlar,~D.~G.; Gagliardi,~L. A new mixing of nonlocal exchange and nonlocal correlation with multiconfiguration pair-density functional theory. \emph{The Journal of Physical Chemistry Letters} \textbf{2020}, \emph{11}, 10158--10163\relax
\mciteBstWouldAddEndPuncttrue
\mciteSetBstMidEndSepPunct{\mcitedefaultmidpunct}
{\mcitedefaultendpunct}{\mcitedefaultseppunct}\relax
\EndOfBibitem
\bibitem[Mostafanejad \latin{et~al.}(2020)Mostafanejad, Liebenthal, and DePrince~III]{mostafanejad2020global}
Mostafanejad,~M.; Liebenthal,~M.~D.; DePrince~III,~A.~E. Global hybrid multiconfiguration pair-density functional theory. \emph{Journal of Chemical Theory and Computation} \textbf{2020}, \emph{16}, 2274--2283\relax
\mciteBstWouldAddEndPuncttrue
\mciteSetBstMidEndSepPunct{\mcitedefaultmidpunct}
{\mcitedefaultendpunct}{\mcitedefaultseppunct}\relax
\EndOfBibitem
\bibitem[Bao \latin{et~al.}(2018)Bao, Gagliardi, and Truhlar]{bao2018self}
Bao,~J.~L.; Gagliardi,~L.; Truhlar,~D.~G. Self-interaction error in density functional theory: An appraisal. \emph{The Journal of Physical Chemistry Letters} \textbf{2018}, \emph{9}, 2353--2358\relax
\mciteBstWouldAddEndPuncttrue
\mciteSetBstMidEndSepPunct{\mcitedefaultmidpunct}
{\mcitedefaultendpunct}{\mcitedefaultseppunct}\relax
\EndOfBibitem
\bibitem[McGuire(1969)]{PhysRev.185.1}
McGuire,~E.~J. $K$-Shell Auger Transition Rates and Fluorescence Yields for Elements Be-Ar. \emph{Physical Review} \textbf{1969}, \emph{185}, 1--6\relax
\mciteBstWouldAddEndPuncttrue
\mciteSetBstMidEndSepPunct{\mcitedefaultmidpunct}
{\mcitedefaultendpunct}{\mcitedefaultseppunct}\relax
\EndOfBibitem
\bibitem[Walters and Bhalla(1971)Walters, and Bhalla]{WALTERS1971301}
Walters,~D.; Bhalla,~C. Nonrelativistic K-shell auger rates and matrix elements for $4 < z \le 54$. \emph{Atomic Data and Nuclear Data Tables} \textbf{1971}, \emph{3}, 301--315\relax
\mciteBstWouldAddEndPuncttrue
\mciteSetBstMidEndSepPunct{\mcitedefaultmidpunct}
{\mcitedefaultendpunct}{\mcitedefaultseppunct}\relax
\EndOfBibitem
\bibitem[Chen \latin{et~al.}(1990)Chen, Larkins, and Crasemann]{CHEN19901}
Chen,~M.~H.; Larkins,~F.~P.; Crasemann,~B. Auger and Coster-Kronig radial matrix elements for atomic numbers $6 \le Z \le 92$. \emph{Atomic Data and Nuclear Data Tables} \textbf{1990}, \emph{45}, 1--205\relax
\mciteBstWouldAddEndPuncttrue
\mciteSetBstMidEndSepPunct{\mcitedefaultmidpunct}
{\mcitedefaultendpunct}{\mcitedefaultseppunct}\relax
\EndOfBibitem
\bibitem[Johnson(1999)]{147901}
Johnson,~R. NIST 101. Computational Chemistry Comparison and Benchmark Database. 1999\relax
\mciteBstWouldAddEndPuncttrue
\mciteSetBstMidEndSepPunct{\mcitedefaultmidpunct}
{\mcitedefaultendpunct}{\mcitedefaultseppunct}\relax
\EndOfBibitem
\bibitem[Wolf \latin{et~al.}(2002)Wolf, Reiher, and Hess]{wolf2002generalized}
Wolf,~A.; Reiher,~M.; Hess,~B.~A. The generalized douglas--kroll transformation. \emph{The Journal of Chemical Physics} \textbf{2002}, \emph{117}, 9215--9226\relax
\mciteBstWouldAddEndPuncttrue
\mciteSetBstMidEndSepPunct{\mcitedefaultmidpunct}
{\mcitedefaultendpunct}{\mcitedefaultseppunct}\relax
\EndOfBibitem
\bibitem[Reiher and Wolf(2004)Reiher, and Wolf]{reiher2004exact}
Reiher,~M.; Wolf,~A. Exact decoupling of the Dirac Hamiltonian. I. General theory. \emph{The Journal of Chemical Physics} \textbf{2004}, \emph{121}, 2037--2047\relax
\mciteBstWouldAddEndPuncttrue
\mciteSetBstMidEndSepPunct{\mcitedefaultmidpunct}
{\mcitedefaultendpunct}{\mcitedefaultseppunct}\relax
\EndOfBibitem
\bibitem[Roos \latin{et~al.}(2004)Roos, Veryazov, and Widmark]{Roos2004}
Roos,~B.~O.; Veryazov,~V.; Widmark,~P.-O. Relativistic atomic natural orbital type basis sets for the alkaline and alkaline-earth atoms applied to the ground-state potentials for the corresponding dimers. \emph{Theoretical Chemistry Accounts} \textbf{2004}, \emph{111}, 345--351\relax
\mciteBstWouldAddEndPuncttrue
\mciteSetBstMidEndSepPunct{\mcitedefaultmidpunct}
{\mcitedefaultendpunct}{\mcitedefaultseppunct}\relax
\EndOfBibitem
\bibitem[Pipek and Mezey(1989)Pipek, and Mezey]{10.1063/1.456588}
Pipek,~J.; Mezey,~P.~G. {A fast intrinsic localization procedure applicable for ab initio and semiempirical linear combination of atomic orbital wave functions}. \emph{The Journal of Chemical Physics} \textbf{1989}, \emph{90}, 4916--4926\relax
\mciteBstWouldAddEndPuncttrue
\mciteSetBstMidEndSepPunct{\mcitedefaultmidpunct}
{\mcitedefaultendpunct}{\mcitedefaultseppunct}\relax
\EndOfBibitem
\bibitem[Aquilante \latin{et~al.}(2007)Aquilante, Lindh, and Bondo~Pedersen]{aquilante2007unbiased}
Aquilante,~F.; Lindh,~R.; Bondo~Pedersen,~T. Unbiased auxiliary basis sets for accurate two-electron integral approximations. \emph{The Journal of Chemical Physics} \textbf{2007}, \emph{127}\relax
\mciteBstWouldAddEndPuncttrue
\mciteSetBstMidEndSepPunct{\mcitedefaultmidpunct}
{\mcitedefaultendpunct}{\mcitedefaultseppunct}\relax
\EndOfBibitem
\bibitem[Gerlach \latin{et~al.}(2022)Gerlach, Preitschopf, Karaev, Quitián-Lara, Mayer, Bozek, Fischer, and Fink]{D2CP02104H}
Gerlach,~M.; Preitschopf,~T.; Karaev,~E.; Quitián-Lara,~H.~M.; Mayer,~D.; Bozek,~J.; Fischer,~I.; Fink,~R.~F. Auger electron spectroscopy of fulminic acid{,} HCNO: an experimental and theoretical study. \emph{Physical Chemistry Chemical Physics} \textbf{2022}, \emph{24}, 15217--15229\relax
\mciteBstWouldAddEndPuncttrue
\mciteSetBstMidEndSepPunct{\mcitedefaultmidpunct}
{\mcitedefaultendpunct}{\mcitedefaultseppunct}\relax
\EndOfBibitem
\bibitem[Schaffner \latin{et~al.}(2024)Schaffner, Gerlach, Karaev, Bozek, Fischer, and Fink]{D4CP03104K}
Schaffner,~D.; Gerlach,~M.; Karaev,~E.; Bozek,~J.; Fischer,~I.; Fink,~R.~F. Experimental and theoretical investigation of the Auger electron spectra of isothiocyanic acid{,} HNCS. \emph{Physical Chemistry Chemical Physics} \textbf{2024}, \emph{26}, 27972--27987\relax
\mciteBstWouldAddEndPuncttrue
\mciteSetBstMidEndSepPunct{\mcitedefaultmidpunct}
{\mcitedefaultendpunct}{\mcitedefaultseppunct}\relax
\EndOfBibitem
\bibitem[Cederbaum and Tarantelli(1993)Cederbaum, and Tarantelli]{10.1063/1.464348}
Cederbaum,~L.~S.; Tarantelli,~F. Nuclear dynamics of decaying states: A time‐dependent formulation. \emph{The Journal of Chemical Physics} \textbf{1993}, \emph{98}, 9691--9706\relax
\mciteBstWouldAddEndPuncttrue
\mciteSetBstMidEndSepPunct{\mcitedefaultmidpunct}
{\mcitedefaultendpunct}{\mcitedefaultseppunct}\relax
\EndOfBibitem
\bibitem[Van~der Straten \latin{et~al.}(1988)Van~der Straten, Morgenstern, and Niehaus]{van1988angular}
Van~der Straten,~P.; Morgenstern,~R.; Niehaus,~A. Angular dependent post-collision interaction in {A}uger processes. \emph{Zeitschrift für Physik D Atoms, Molecules and Cluster} \textbf{1988}, \emph{8}, 35--45\relax
\mciteBstWouldAddEndPuncttrue
\mciteSetBstMidEndSepPunct{\mcitedefaultmidpunct}
{\mcitedefaultendpunct}{\mcitedefaultseppunct}\relax
\EndOfBibitem
\bibitem[Rohatgi(2024)]{WebPlotDigitizer}
Rohatgi,~A. WebPlotDigitizer. \url{https://github.com/automeris-io/WebPlotDigitizer}, 2024; Version 4.7. Licensed under GNU AGPL v3.\relax
\mciteBstWouldAddEndPunctfalse
\mciteSetBstMidEndSepPunct{\mcitedefaultmidpunct}
{}{\mcitedefaultseppunct}\relax
\EndOfBibitem
\bibitem[Correia \latin{et~al.}(1991)Correia, Naves~de Brito, Keane, Karlsson, Svensson, Liegener, Cesar, and Ågren]{10.1063/1.461687}
Correia,~N.; Naves~de Brito,~A.; Keane,~M.~P.; Karlsson,~L.; Svensson,~S.; Liegener,~C.; Cesar,~A.; Ågren,~H. {Doubly charged valence states of formaldehyde, acetaldehyde, acetone, and formamide studied by means of photon excited Auger electron spectroscopy and ab initio calculations}. \emph{The Journal of Chemical Physics} \textbf{1991}, \emph{95}, 5187--5197\relax
\mciteBstWouldAddEndPuncttrue
\mciteSetBstMidEndSepPunct{\mcitedefaultmidpunct}
{\mcitedefaultendpunct}{\mcitedefaultseppunct}\relax
\EndOfBibitem
\bibitem[Houston and Rye(1981)Houston, and Rye]{10.1063/1.440831}
Houston,~J.~E.; Rye,~R.~R. {Auger electron spectra of the cycloalkanes C3 through C6}. \emph{The Journal of Chemical Physics} \textbf{1981}, \emph{74}, 71--76\relax
\mciteBstWouldAddEndPuncttrue
\mciteSetBstMidEndSepPunct{\mcitedefaultmidpunct}
{\mcitedefaultendpunct}{\mcitedefaultseppunct}\relax
\EndOfBibitem
\bibitem[Rye \latin{et~al.}(1980)Rye, Jennison, and Houston]{10.1063/1.440015}
Rye,~R.~R.; Jennison,~D.~R.; Houston,~J.~E. {Auger spectra of alkanes}. \emph{The Journal of Chemical Physics} \textbf{1980}, \emph{73}, 4867--4874\relax
\mciteBstWouldAddEndPuncttrue
\mciteSetBstMidEndSepPunct{\mcitedefaultmidpunct}
{\mcitedefaultendpunct}{\mcitedefaultseppunct}\relax
\EndOfBibitem
\bibitem[Schnack-Petersen \latin{et~al.}(2022)Schnack-Petersen, Tenorio, Coriani, Decleva, Troß, Ramasesha, Coreno, Totani, and Röder]{10.1063/5.0122088}
Schnack-Petersen,~A.~K.; Tenorio,~B. N.~C.; Coriani,~S.; Decleva,~P.; Troß,~J.; Ramasesha,~K.; Coreno,~M.; Totani,~R.; Röder,~A. {Core spectroscopy of oxazole}. \emph{The Journal of Chemical Physics} \textbf{2022}, \emph{157}, 214305\relax
\mciteBstWouldAddEndPuncttrue
\mciteSetBstMidEndSepPunct{\mcitedefaultmidpunct}
{\mcitedefaultendpunct}{\mcitedefaultseppunct}\relax
\EndOfBibitem
\bibitem[Neeb \latin{et~al.}(1997)Neeb, Kivimäki, Kempgens, Köppe, and Bradshaw]{MNeeb_1997}
Neeb,~M.; Kivimäki,~A.; Kempgens,~B.; Köppe,~H.~M.; Bradshaw,~A.~M. The C 1s Auger decay spectrum of CF4: an analysis of the core-excited states. \emph{Journal of Physics B: Atomic, Molecular and Optical Physics} \textbf{1997}, \emph{30}, 93\relax
\mciteBstWouldAddEndPuncttrue
\mciteSetBstMidEndSepPunct{\mcitedefaultmidpunct}
{\mcitedefaultendpunct}{\mcitedefaultseppunct}\relax
\EndOfBibitem
\bibitem[Suzuki \latin{et~al.}(2019)Suzuki, Hino, Kotsugi, and Ono]{Suzuki2019}
Suzuki,~Y.; Hino,~H.; Kotsugi,~M.; Ono,~K. Automated estimation of materials parameter from X-ray absorption and electron energy-loss spectra with similarity measures. \emph{npj Computational Materials} \textbf{2019}, \emph{5}, 39\relax
\mciteBstWouldAddEndPuncttrue
\mciteSetBstMidEndSepPunct{\mcitedefaultmidpunct}
{\mcitedefaultendpunct}{\mcitedefaultseppunct}\relax
\EndOfBibitem
\bibitem[Chen \latin{et~al.}(2024)Chen, Chen, Hwang, Davis, Yang, Sun, Lee, McReynolds, Allan, Marulanda~Arias, Ong, and Chan]{doi:10.1021/acs.chemmater.3c02584}
Chen,~Y.; Chen,~C.; Hwang,~I.; Davis,~M.~J.; Yang,~W.; Sun,~C.; Lee,~G.-H.; McReynolds,~D.; Allan,~D.; Marulanda~Arias,~J.; Ong,~S.~P.; Chan,~M. K.~Y. Robust Machine Learning Inference from X-ray Absorption Near Edge Spectra through Featurization. \emph{Chemistry of Materials} \textbf{2024}, \emph{36}, 2304--2313\relax
\mciteBstWouldAddEndPuncttrue
\mciteSetBstMidEndSepPunct{\mcitedefaultmidpunct}
{\mcitedefaultendpunct}{\mcitedefaultseppunct}\relax
\EndOfBibitem
\bibitem[Sen and Ghosh(2024)Sen, and Ghosh]{doi:10.1021/acs.jpcc.4c01750}
Sen,~A.; Ghosh,~S. Understanding the Empirical Shifts Required for Quantitative Computation of X-ray Spectroscopy. \emph{The Journal of Physical Chemistry C} \textbf{2024}, \emph{128}, 10871--10879\relax
\mciteBstWouldAddEndPuncttrue
\mciteSetBstMidEndSepPunct{\mcitedefaultmidpunct}
{\mcitedefaultendpunct}{\mcitedefaultseppunct}\relax
\EndOfBibitem
\bibitem[Fouda and Besley(2017)Fouda, and Besley]{Fouda2017}
Fouda,~A. E.~A.; Besley,~N.~A. Assessment of basis sets for density functional theory-based calculations of core-electron spectroscopies. \emph{Theoretical Chemistry Accounts} \textbf{2017}, \emph{137}, 6\relax
\mciteBstWouldAddEndPuncttrue
\mciteSetBstMidEndSepPunct{\mcitedefaultmidpunct}
{\mcitedefaultendpunct}{\mcitedefaultseppunct}\relax
\EndOfBibitem
\bibitem[Carlson \latin{et~al.}(2015)Carlson, Truhlar, and Gagliardi]{doi:10.1021/acs.jctc.5b00609}
Carlson,~R.~K.; Truhlar,~D.~G.; Gagliardi,~L. Multiconfiguration Pair-Density Functional Theory: A Fully Translated Gradient Approximation and Its Performance for Transition Metal Dimers and the Spectroscopy of Re2Cl82–. \emph{Journal of Chemical Theory and Computation} \textbf{2015}, \emph{11}, 4077--4085, PMID: 26575903\relax
\mciteBstWouldAddEndPuncttrue
\mciteSetBstMidEndSepPunct{\mcitedefaultmidpunct}
{\mcitedefaultendpunct}{\mcitedefaultseppunct}\relax
\EndOfBibitem
\bibitem[Rennie \latin{et~al.}(2000)Rennie, Kempgens, Köppe, Hergenhahn, Feldhaus, Itchkawitz, Kilcoyne, Kivimäki, Maier, Piancastelli, Polcik, Rüdel, and Bradshaw]{10.1063/1.1290029}
Rennie,~E.~E.; Kempgens,~B.; Köppe,~H.~M.; Hergenhahn,~U.; Feldhaus,~J.; Itchkawitz,~B.~S.; Kilcoyne,~A. L.~D.; Kivimäki,~A.; Maier,~K.; Piancastelli,~M.~N.; Polcik,~M.; Rüdel,~A.; Bradshaw,~A.~M. {A comprehensive photoabsorption, photoionization, and shake-up excitation study of the C 1s cross section of benzene}. \emph{The Journal of Chemical Physics} \textbf{2000}, \emph{113}, 7362--7375\relax
\mciteBstWouldAddEndPuncttrue
\mciteSetBstMidEndSepPunct{\mcitedefaultmidpunct}
{\mcitedefaultendpunct}{\mcitedefaultseppunct}\relax
\EndOfBibitem
\bibitem[Hitchcock \latin{et~al.}(1986)Hitchcock, Newbury, Ishii, Stöhr, Horsley, Redwing, Johnson, and Sette]{10.1063/1.451719}
Hitchcock,~A.~P.; Newbury,~D.~C.; Ishii,~I.; Stöhr,~J.; Horsley,~J.~A.; Redwing,~R.~D.; Johnson,~A.~L.; Sette,~F. {Carbon K‐shell excitation of gaseous and condensed cyclic hydrocarbons: C3H6, C4H8, C5H8, C5H1, C6H1, C6H12, and C8H8}. \emph{The Journal of Chemical Physics} \textbf{1986}, \emph{85}, 4849--4862\relax
\mciteBstWouldAddEndPuncttrue
\mciteSetBstMidEndSepPunct{\mcitedefaultmidpunct}
{\mcitedefaultendpunct}{\mcitedefaultseppunct}\relax
\EndOfBibitem
\bibitem[Pireaux \latin{et~al.}(1976)Pireaux, Svensson, Basilier, Malmqvist, Gelius, Caudano, and Siegbahn]{PhysRevA.14.2133}
Pireaux,~J.~J.; Svensson,~S.; Basilier,~E.; Malmqvist,~P.-A.; Gelius,~U.; Caudano,~R.; Siegbahn,~K. Core-electron relaxation energies and valence-band formation of linear alkanes studied in the gas phase by means of electron spectroscopy. \emph{Physical Review A} \textbf{1976}, \emph{14}, 2133--2145\relax
\mciteBstWouldAddEndPuncttrue
\mciteSetBstMidEndSepPunct{\mcitedefaultmidpunct}
{\mcitedefaultendpunct}{\mcitedefaultseppunct}\relax
\EndOfBibitem
\bibitem[Plekan \latin{et~al.}(2020)Plekan, Sa’adeh, Ciavardini, Callegari, Cautero, Dri, Di~Fraia, Prince, Richter, Sergo, Stebel, Devetta, Faccialà, Vozzi, Avaldi, Bolognesi, Castrovilli, Catone, Coreno, Zuccaro, Bernes, Fronzoni, Toffoli, and Ponzi]{doi:10.1021/acs.jpca.0c02719}
Plekan,~O. \latin{et~al.}  Experimental and Theoretical Photoemission Study of Indole and Its Derivatives in the Gas Phase. \emph{The Journal of Physical Chemistry A} \textbf{2020}, \emph{124}, 4115--4127\relax
\mciteBstWouldAddEndPuncttrue
\mciteSetBstMidEndSepPunct{\mcitedefaultmidpunct}
{\mcitedefaultendpunct}{\mcitedefaultseppunct}\relax
\EndOfBibitem
\bibitem[Storchi \latin{et~al.}(2008)Storchi, Tarantelli, Veronesi, Bolognesi, Fainelli, and Avaldi]{10.1063/1.2993317}
Storchi,~L.; Tarantelli,~F.; Veronesi,~S.; Bolognesi,~P.; Fainelli,~E.; Avaldi,~L. The Auger spectroscopy of pyrimidine and halogen-substituted pyrimidines. \emph{The Journal of Chemical Physics} \textbf{2008}, \emph{129}, 154309\relax
\mciteBstWouldAddEndPuncttrue
\mciteSetBstMidEndSepPunct{\mcitedefaultmidpunct}
{\mcitedefaultendpunct}{\mcitedefaultseppunct}\relax
\EndOfBibitem
\bibitem[Jangid \latin{et~al.}()Jangid, Hermes, and Gagliardi]{jangid2024core}
Jangid,~B.; Hermes,~M.~R.; Gagliardi,~L. Core Binding Energy Calculations: A Scalable Approach with the Quantum Embedding-Based Equation-of-Motion Coupled-Cluster Method. \emph{The Journal of Physical Chemistry Letters} \emph{15}, 5954--5963\relax
\mciteBstWouldAddEndPuncttrue
\mciteSetBstMidEndSepPunct{\mcitedefaultmidpunct}
{\mcitedefaultendpunct}{\mcitedefaultseppunct}\relax
\EndOfBibitem
\end{mcitethebibliography}

\end{document}